\documentclass[11pt,a4paper]{article}
\usepackage{jheppub}
\usepackage{amsmath,amssymb,amsfonts,mathtools}
\usepackage{epsfig}
\usepackage{graphicx}
\usepackage{color}
\usepackage{shuffle}
\usepackage{hyperref}
\usepackage{cleveref}
\usepackage[dvipsnames]{xcolor}
\usepackage{graphicx}
\usepackage{hhline}
\usepackage{contour}
\usepackage[tableposition=above]{caption}
\usepackage[normalem]{ulem}
\usepackage{tikz-feynman} 
\tikzfeynmanset{compat=1.1.0}
\tikzfeynmanset{/tikzfeynman/warn luatex = false}
\usetikzlibrary{backgrounds,calc}
\usepackage{wrapfig}
\usepackage{cases}
\usepackage{diagbox}
\usepackage{booktabs}
\usepackage{orcidlink}

\usepackage{tikz}
\usetikzlibrary{positioning, shapes.geometric, arrows.meta,calc}
\tikzset{
  optional arrow/.style={
    -{Stealth[length=3mm]},
    thick,
    dashed,
    gray!60
  },
  optional block/.style={
    draw=gray!90,
    fill=cyan!20  
  }
}

\usepackage{slashed}
\usepackage{caption}
\usepackage{subcaption}
\usepackage{array}

\DeclareMathOperator{\tr}{tr}

\newcommand{\pol}{\ensuremath{\epsilon}}

\newcommand{\spinBlock}[3]{\ensuremath{n_{#1}^{(#2|#3)}}}
\newcommand{\spinBlockS}[3]{\spinBlock{2s,#1}{#2}{#3}}
\newcommand{\spinBlockF}[3]{\spinBlock{4\psi,#1}{#2}{#3}}
\newcommand{\spinBlockV}[3]{\spinBlock{2v,#1}{#2}{#3}}
\newcommand{\spinBlockSV}[2]{\ensuremath{n_{sv,#1}^{(#2)}}}

\newcommand{\vfBlock}[1]{\ensuremath{n_{#1}}}

\newcommand{\pbarp}{\ensuremath{\overline{\psi}_1 \psi_2}}
\newcommand{\pbarpexpr}[1]{\ensuremath{\overline{\psi}_1 #1 \psi_2}}
\newcommand{\pbarpe}[1]{\pbarpexpr{\slashed{\pol}_{#1}}}
\newcommand{\pbarpk}[1]{\pbarpexpr{\slashed{k}_{#1}}}
\newcommand{\pbarpf}[1]{\ensuremath{F_{#1}^{\mu}{}_{\nu}\, \pbarpexpr{\gamma^\nu}}}
\newcommand{\pbarpDoub}{\pbarpexpr{\gamma^{\mu \nu} \psi_2}}
\newcommand{\pbarpTrip}{\pbarpexpr{\gamma^{\mu \nu \rho}}}

\newcommand{\kslashdiff}[2]{\ensuremath{(\slashed{k}_{#1} - \slashed{k}_{#2})}}

\newcommand{\pbarpgenexpr}[3]{\ensuremath{\overline{\psi}_{#1} #3 \psi_{#2}}}

\newcommand{\tzero}{\ensuremath{-}}
\newcommand{\tone}{\ensuremath{+}}

\newcommand{\nc}{\newcommand}
\nc{\la}{\langle}
\nc{\ra}{\rangle}
\nc{\nn}{\nonumber \\}

\newcommand{\ppp}[1]{\ensuremath{\mathcal{P}^{(+|+)}_{#1}}}
\newcommand{\pmm}[1]{\ensuremath{\mathcal{P}^{(-|-)}_{#1}}}
\newcommand{\ppm}[1]{\ensuremath{\mathcal{P}^{(+|-)}_{#1}}}
\newcommand{\pmp}[1]{\ensuremath{\mathcal{P}^{(-|+)}_{#1}}}
\newcommand{\phh}[3]{\ensuremath{\mathcal{P}^{(#1|#2)}_{#3}}}

\newcommand{\chh}[3]{\ensuremath{\mathcal{C}^{(#1|#2)}_{#3}}}
\newcommand{\cpp}[1]{\chh{+}{+}{#1}}
\newcommand{\cmm}[1]{\chh{-}{-}{#1}}

\newcommand{\NA}{{\text{NA}}} 

\newcommand{\cnn}[1]{\ensuremath{\mathcal{C}^{( \NA | \NA)}_{#1}}}
\newcommand{\cpn}[1]{\ensuremath{\mathcal{C}^{(+| \NA) }_{#1}}}
\newcommand{\cmn}[1]{\ensuremath{\mathcal{C}^{(-| \NA )}_{#1}}}
\newcommand{\cnm}[1]{\ensuremath{\mathcal{C}^{(\NA |-)}_{#1}}}
\newcommand{\cnp}[1]{\ensuremath{\mathcal{C}^{(\NA |+)}_{#1}}}

\newcommand{\chhfull}[3]{\ensuremath{\mathfrak{C}_{#3}^{(#1|#2)}}}

\newcommand{\dherm}{\ensuremath{\overleftrightarrow{D}}}

\usepackage[utf8]{inputenc}
\DeclareUnicodeCharacter{03BE}{\xi}
\DeclareUnicodeCharacter{03D5}{\phi}

\newcommand{\ie}{i.e.~}
\newcommand{\eg}{e.g.~}

\title{$D$-Dimensional Modular Assembly of Higher-Derivative Four-Point Contact Amplitudes Involving Fermions}
\author[a,b]{John Joseph M. Carrasco\orcidlink{0000-0002-4499-8488},}
\author[a]{Sai Sasank Chava\orcidlink{0009-0009-0228-9737},}
\author[a,b]{Alex Edison\orcidlink{0000-0002-5430-9500},}
\author[a]{Aslan Seifi\orcidlink{0009-0007-2322-5842}}

\affiliation[a]{The Amplitudes and Insights Group, Department of Physics \& Astronomy, Northwestern University, Evanston, Illinois 60208, USA}
\affiliation[b]{Center for Interdisciplinary Exploration and Research in Astrophysics (CIERA), Northwestern University, 1800 Sherman Ave, Evanston, IL 60201, USA}

\emailAdd{carrasco@northwestern.edu}
\emailAdd{chava@u.northwestern.edu}
\emailAdd{alexander.edison@northwestern.edu}
\emailAdd{AslanSeifi2024@u.northwestern.edu}

\date{\today}

\abstract{
    We present a novel robust framework for systematically constructing
    $D$-dimensional four-point higher-derivative contact amplitudes. Our modular
    block (``LEGO''-like) approach builds amplitudes directly from manifestly
    gauge-invariant kinematic blocks, color-weight factors, and scalar
    Mandelstam polynomials. Symmetries (Bose/Fermi) are imposed algebraically,
    acting as filters on combinations of compatible pieces. This framework
    operates entirely in $D$ dimensions, naturally incorporating evanescent
    operators crucial for loop-level consistency. Scaling to arbitrary mass
    dimension is achieved in a highly controlled manner using
    permutation-invariant scalar polynomials, avoiding combinatorial explosion.
    A key feature is its manifest compatibility with the double-copy program,
    allowing the systematic generation of operator towers not only for gauge
    theories but also for gravity and other theories within the double-copy web.
}         

\begin{document} 

\maketitle
\flushbottom

\section{Introduction}
\label{sec:Introduction}

The systematic construction of higher-derivative interactions in quantum field
theories is essential for robust Effective Field Theory (EFT) frameworks.
Traditional approaches, however,  confront  significant challenges. Ensuring a
complete, non-redundant basis of operators that respects all gauge and exchange
symmetries is a critical part of the
process~\cite{Lehman:2014jma,Lehman:2015coa,Liao:2020jmn,Li:2020xlh,Li:2020gnx,Ma:2019gtx,Murphy:2020rsh,Christensen:2018zcq}.  This often involves intricate
multi-stage procedures, especially when operating in $D$-dimensions is required
to correctly capture loop-level effects and to enable the construction of consistent loop integrands.  
The more moving parts to a procedure,
the more opportunities for redundant descriptions and potential inconsistencies. 

In this work, we are guided by the engineering principle of keeping it simple.
We introduce, for interactions involving fermions, a novel and systematic LEGO
bootstrap method for constructing four-point higher-derivative contact
amplitudes out of manageable, individually self-consistent pieces.  These are designed to serve as modular building blocks for both tree-level EFT matching and loop-integrand assembly, either by mapping back to local operators and using Feynman rules, or directly through on-shell, $D$-dimensional unitarity-based constructions.
LEGO here
stands for Local Effective Gauge Operators, and evokes the modular building
block toy in their ability to snap together in well-defined ways. Our approach
begins by identifying D-dimensional manifestly gauge-invariant kinematic blocks,
fundamental color structures, and structured scalar polynomials of Mandelstam
variables, where each of these blocks has well-defined properties and are
finitely generated. A critical realization enabling this is that all vector
components can always be expressed in terms of linearized field-strengths.  This
simplifies both their interplay with color factors, as well as facilitates
constructing kinematic building blocks with well defined symmetry properties.
With the blocks in hand, we assemble them into amplitudes using Bose/Fermi
statistics as direct compatibility conditions between the blocks. The entire
procedure is performed in general D dimensions to ensure even evanescent
operators are covered.  Notably, these operators are vital for ensuring
consistent loop calculations
\cite{Bern:2017puu,Bern:2017tuc,Bern:2017rjw,Bern:2019isl,Fuentes-Martin:2022vvu}\,.

Furthermore, our method exhibits control and scalability to arbitrarily high
mass dimensions.  The potential combinatorial explosion of terms is kept under
control by organizing the scalar Mandelstam dependence according to the required
permutation properties. This ensures that increasing the derivative order
primarily involves adjusting powers of these invariants, rather than introducing
fundamentally new complex structures.

It is important to note that in this work we restrict attention to spacetime-parity–even operators, which admit a clean and unambiguous formulation in arbitrary spacetime dimension. Parity-odd structures involving $\gamma_5$ or Levi-Civita tensors necessarily require dimension-specific prescriptions which we look forward to addressing in future work..

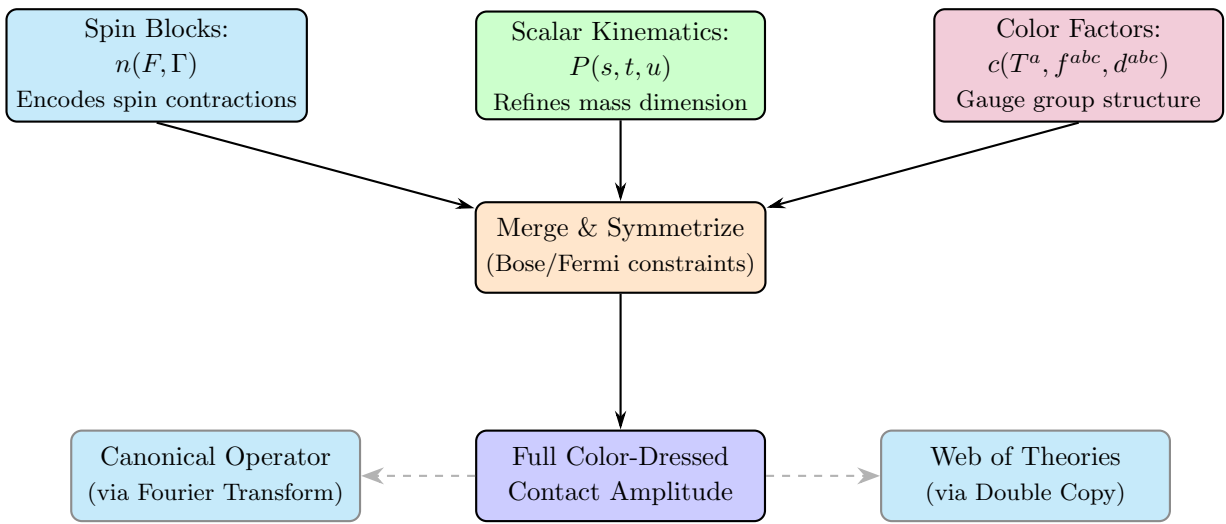
\begin{figure}[htbp] 
\centering
\begin{tikzpicture}[
  node distance=1.8cm and 2.2cm, 
  every node/.style={align=center, font=\small},
  block/.style={draw, rectangle, rounded corners, thick, minimum width=3.8cm, minimum height=1.2cm}, 
  arrow/.style={-{Stealth[length=2.5mm, width=1.5mm]}, thick} 
]

  \node[block, fill=cyan!20] (spin) {Spin Blocks: \\$n(F, \Gamma)$\\ {\footnotesize Encodes spin contractions}};
  \node[block, fill=green!20, right=of spin] (poly) {Scalar Kinematics:\\$P(s,t,u)$\\ {\footnotesize Refines mass dimension}}; 
  \node[block, fill=purple!20, right=of poly] (color) {Color Factors:\\ $c(T^a, f^{abc},d^{abc})$\\ {\footnotesize Gauge group structure}}; 
  
  \node[block, fill=orange!20, below=of $(poly)$ 
    ] (merge) {Merge \& Symmetrize\\ {\footnotesize (Bose/Fermi constraints)}}; 

  \node[block, fill=blue!20, below=of merge] (full) {Full Color‑Dressed\\Contact Amplitude};
  
  \node[block, optional block, right=1.5cm of full] (grav) 
  {{Web of Theories}\\{\footnotesize {(via Double Copy)}} };
  \node[block, optional block, left=1.5cm of full] (op) 
    {{Canonical Operator}\\{\footnotesize {(via Fourier Transform)}} };

  \draw[arrow] (spin.south) -- (merge);
  \draw[arrow] (poly.south) -- (merge);
  \draw[arrow] (color.south) -- (merge);
  
  \draw[arrow] (merge.south) -- (full.north);
  
  \draw[optional arrow] (full.east) -- (grav.west);
    \draw[optional arrow] (full.west) -- (op.east);

\end{tikzpicture}
\caption{Schematic workflow for the ``LEGO-like'' modular bootstrap of four-point
    contact amplitudes. Spin-dependent blocks (from $F, \Gamma$), scalar
    kinematic polynomials (in $s,t,u$), and color factors are combined according
    to well-defined compatibility. Symmetry constraints (Bose/Fermi) are applied
    during the merge step to yield the full D-dimensional, gauge-invariant
    amplitude. An optional subsequent step (dashed) can relate these to
    predictions in a wide web of EFT theories via double copy. It is
    straightforward to encode these contact structures at the operator level having
    identified the distinct predictive building blocks.
}
\label{fig:lego_bootstrap_workflow} 
\end{figure}

We present the schematic workflow of our modular approach in
\cref{fig:lego_bootstrap_workflow}.  The spin blocks $n(\Gamma,F)$ capture all
spin-dependent contractions and manifest Lorentz invariance.   The final mass
dimension of the operator is refined via scalar polynomials of Mandelstam
invariants in $P(s,t,u)$.  The gauge group structure is encoded in the color
factors $c(T^a, f^{abc},d^{abc})$.  These are combined based upon compatibility
criteria depending on desired field properties like parity under Bose and Fermi
exchange, resulting in the full color-dressed amplitude.  While the procedure to
encode arbitrary multiplicity amplitude structures at the operator level will
be described elsewhere, at four-points the structures are actually quite simple.
We will describe the straightforward mapping to canonical operators for these
specific contact terms, after which Wilson coefficients can be assigned to the
basis.  Due to the sharply defined symmetry properties, and
manifest gauge invariance, these amplitudes play well with and define
contributions to the wide web of double-copy theories~\cite{Kawai:1985xq,Bern:2008qj,Bern:2010ue,Cachazo:2013iea,Cachazo:2013hca,Bern:2019prr}.

\subsection{Relation to prior work}
\label{sec:review}

The study of Standard Model Effective Field Theory (SMEFT) operators beyond
dimension-six is crucial for interpreting precision measurements and searching
for new physics. Theoretical developments in this area, including operator
enumeration, construction, renormalization, and phenomenological impact,
particularly at dimension-8 and beyond, have been recently reviewed in the
Snowmass 2021 process~\cite{Alioli:2022fng}.

We highlight a few developments of particular relevance to this work.  Much of
the recent development in SMEFT operator enumeration has been predicated on a
robust understanding of \emph{how many} operators are supported by the Standard
Model symmetries, for any given particle configuration.  The application of
invariant theory methods to this task has been foundational to the
program~\cite{Henning:2017fpj, Hays:2018zze, Henning:2019mcv, Henning:2019enq}.
In parallel, developments of on-shell spinor helicity methods for massive states
\cite{Arkani-Hamed:2017jhn} has led to systematic construction of three and
four-point contact amplitudes in four spacetime dimensions
\cite{Durieux:2020gip}. Such little-group covariant formalisms allow for
incredibly compact expressions given fixed external states, especially for
contact terms, and gives a handle on bookkeeping that can be exploited to track
gauge and flavor symmetries. Notably, the 'Young Tensor'
approach~\cite{Li:2020gnx,Li:2020zfq,Li:2022tec,Li:2020xlh,Harlander:2023psl} established a
framework for systematically mapping little-group properties for the purpose of
operator enumeration. This approach has been implemented in the comprehensive
software packages ABC4EFT~\cite{Li:2022tec} and AutoEFT~\cite{Harlander:2023ozs}, which aim to provide complete
N-point operator bases for generic EFTs, including the full SM field content.

Our work presents a $D$-dimensional modular approach for constructing
higher-derivative four-point contact interactions. While sharing the
foundational principle from spinor-helicity methods of building from
gauge-invariant blocks, our approach differs in its explicit D-dimensional
construction of the kinematic blocks from the outset. This inherently
incorporates evanescent operator structures directly into the basis, which is a
genuinely important feature for the consistency of loop-induced effective
operators and the investigation of finite
counterterms (see,
e.g.~refs.~\cite{Bern:2017puu,Bern:2017tuc,Bern:2017rjw,Bern:2019isl,Fuentes-Martin:2022vvu}) 
especially when massless vectors (or gravitons) are involved. 

We are at the beginning of the exploration.  Three point amplitudes are trivial
(see \cref{app:3pt}), but the
four-points success presented here represents an important proof of concept. Our
exploration has allowed us to demonstrate a distinct factorization into
[Spin]$\times$ [Color] $\times$ [Scalar].  Notably the scalar polynomials, built from
$D$-dimensional Mandelstam invariants (e.g., permutation invariants),
systematically drive the progression to arbitrary mass dimension, while the
color and spin blocks have finite and manageable bases. The modular approach we
present here builds upon prior
work~\cite{Carrasco:2019yyn,Carrasco:2021ptp,Carrasco:2022lbm,Carrasco:2022jxn,Carrasco:2023wib}
understanding color-dual Yang-Mills amplitudes to all order corrections in the
UV at four points and five points, as well as recent work by two of the current
authors functionalizing fundamental fermion dressings towards color-dual
loop-level QCD~\cite{Carrasco:2023vjg}.

\subsection{Roadmap}
  The structure of our main body will follow the modular building
blocks as described: scalar blocks in \cref{sec:scalars}, followed by the color
blocks in \cref{sec:color}, and ending with the spin blocks in
\cref{sec:spinors}.
Then, we describe the merging procedure for building ``complete''
amplitudes in \cref{sec:merge}, ease of double-copy in \cref{sec:DC},
 and provide a number of examples in \cref{sec:examples}.
Special considerations in four dimensions are described in  
\cref{sec:4F4D,sec:sh-proj}.
We provide an ancillary machine-readable data file with the arXiv preprint for the spin blocks involving
bosons.

\section{Scalar building blocks and Kinematics}
\label{sec:scalars}
\subsection{Leg Godt: Scalar kinematics definite-parity blocks}
\label{sec:slego}
Any arbitrary scalar function $f(1\dots n)$ can be decomposed into the sum of components that are symmetric or antisymmetric under the exchange $1 \leftrightarrow 2$:
\begin{align}
    f(123 \dots n) = \frac{f(123 \dots n) + f(213 \dots n)}{2}+\frac{f(123 \dots n) - f(213 \dots n)}{2}.
\end{align}
As such, the space of scalar functions corresponding to four-particle kinematics
can be organized as a direct sum of scalar blocks with definite parity under $1
\leftrightarrow 2$ or $3 \leftrightarrow 4$. The advantage of such a
classification is that the scalar building blocks play well with our other
modular blocks when they are organized into families of discrete parity under
exchange. In this paper, we specifically consider the case where particles $1$
and $2$ belong to one family and $3$ and $4$ belong to another (not necessarily
distinct) family. We represent the space of scalar blocks compatible with parity
$h_1$ and $h_2$ under $1\leftrightarrow 2$ and $ 3 \leftrightarrow 4$,
respectively, at mass dimension $D$ with $\mathcal{P}_D ^{(h_1 | h_2)}$. 

We take Mandelstam definitions as follows:
\begin{align}
    s=(k_1+k_2)^2 \hspace{1cm} t=(k_2+k_3)^2\hspace{1cm} u=(k_1+k_3)^2.
\end{align}
It is useful to define the mass invariants that show up often in our scalar basis. For general masses \(m_1,m_2,m_3,m_4\) we define left/right sums and products
\begin{equation}
\mu_{1,\,l}=m_1+m_2,\quad
\mu_{2,\,l}=m_1m_2,
\qquad
\mu_{1,\,r}=m_3+m_4,\quad
\mu_{2,\,r}=m_3m_4.
\label{eq:gen-masses}
\end{equation}
These functions have definite parity $+$ under $1 \leftrightarrow 2$ or $3 \leftrightarrow
4$. In terms of these combinations, we can write any scalar
function of definite parity and mass-dimension $D$ as follows:
\begin{itemize}
  \item[\((+|+)\)] Even-Even:
  \begin{equation}
\ppp{D} = s^{a_1}\bigl((t-u)^{2}\bigr)^{\, a_2}
\mu_{1,\,l}^{\,a_3}\mu_{1,\,r}^{\,a_4}\mu_{2,\,l}^{\,a_5}\mu_{2,\,r}^{\,a_6}
\bigl((t-u)(m_{1}-m_{2})(m_{3}-m_{4})\bigr)^{a_7} \, ,
\label{eq:slego-pp-4}
\end{equation}
with  \( D = (a_3 + a_4)+2 (a_1 + a_5 + a_6) +4 ( a_2 + a_7) \) and $a_7 \in
\{0,1\}$.

 \item[\((- | +)\)]  Odd-Even:  
    \begin{equation}
    \pmp{D,1}=(m_1-m_2) \ppp{D-1}  \qquad   \pmp{D,2}=(m_3-m_4)(t-u) \ppp{D-3}  \, .
    \label{eq:slego-mp}
    \end{equation}

 \item[\((+ | -)\)] Even-Odd:  
   \begin{equation}
    \ppm{D,1}=(m_3-m_4) \ppp{D-1}\,  \qquad   \ppm{D,2}=(m_1-m_2)(t-u) \ppp{D-3} \, .
    \label{eq:slego-pm}
   \end{equation}

  \item[\((- | -)\)] Odd-Odd:  
\begin{equation}
\pmm{D,1} = (t-u) \ppp{D-2} \qquad
\pmm{D,2} = (m_1-m_2)(m_3-m_4)\ppp{D-2}\\.
\label{eq:slego-mm}
\end{equation}
\end{itemize}
Note that the two types of terms in each of
\cref{eq:slego-mp,eq:slego-pm,eq:slego-mm} are not strictly independent when
they involve terms from $\ppp{D}$ with $a_7=1$.
We will refer to \textit{all} scalar functions at mass dimension $D$ as $\mathcal{P}_D$ in this paper.

\subsection{Special cases}
The basis listed above holds for all arbitrary masses $m_1,m_2,m_3$ and $m_4$.
Many simplifications occur when we set some of the masses equal to the others. We
lay out some of the special cases below. 

\paragraph{Two mass families:}
Consider the case where we have \(m_1=m_2 \equiv m_{f,1}\), \(m_3=m_4 \equiv m_{f,2}\).  
In this case,  the \( (+|+) \) scalar block reduces to
\begin{equation}
\ppp{D} = s^{a_1} \bigl((t-u)^{2}\bigr)^{\, a_2} m_{f_1 }^{\,a_3} m_{f_2}^{\,a_4}, 
\label{eq:slego-pp-2}
\end{equation}
with dimensions satisfying
\begin{align}
D &= (a_3 + a_4)+ 2 ( a_1 ) +4 ( a_2).\,
\end{align}  
The $\ppm{}$ and $\pmp{}$ cases vanish, and only \(\pmm{D,1}\) survives among the odd–odd blocks. The case where particles $3$ and $4$ are massless can be obtained by simply setting $a_4=0$.

\paragraph{Entirely agnostic to mass:} There can be cases either when the
particles are massless, or one may wish to handle mass using alternative
considerations, in which case the appropriate blocks follow from:
\begin{equation}
\ppp{D} = s^{a_1} \left((t-u)^{2}\right)^{\, a_2}\, \qquad \pmm{D} = (t-u) \ppp{D-2}
\label{eq:slego-pp-0}
\end{equation}
with \(D=2 a_1+4 a_2 \) . 

\subsection{Useful Decompositions}
The well-known elementary permutation invariants $s^n + t^n + u^n$ can of course be
decomposed according to \cref{eq:slego-pp-0}:
\begin{align}
    \sigma_2 &= (s^2 + t^2 + u^2)/2 = \frac{3}{4}  \underbrace{s^2}_{\ppp{4+0}}
    +\frac{1}{4} \underbrace{(t-u)^{2}}_{\ppp{0 + 4}}\label{eq:sigma-2}\\
    \sigma_3 &=s\, t\, u = \frac{1}{3} \left(s^3 + t^3 + u^3\right) =
    \frac{1}{4} \left( \underbrace{s^3}_{\ppp{6+0}} -\underbrace{s
    (t-u)^2}_{\ppp{2+4}}  \right)\label{eq:sigma-3}\\
    s^n + t^n + u^n &= \underbrace{s^n}_{\ppp{2 n + 0}}
    +2^{-n}(-1)^{n}
    \sum_{j=0}^{\lfloor n/2\rfloor}
    \binom{n}{2j}\,
    \underbrace{s^{\,n-2j}\,(t-u)^{2j}}_{\ppp{2 (n-2j) + 2(2j)}}\,.
    \label{eq:sigma-n}
\end{align}

\section{Color Blocks}
\label{sec:color}

The \emph{color weights} dictate how gauge charges and flavor quantum numbers
flow through the diagram.  We adopt channel labels according to the connectivity
of a ``fictitious'' internal propagator:
\begin{itemize}
  \item \textbf{s-channel:} legs $(1,2)\to(3,4)$
  \item \textbf{t-channel:} legs $(4,1)\to(2,3)$
  \item \textbf{u-channel:} legs $(3,1)\to(4,2)$
\end{itemize}

Adjoint representations have the usual odd-parity color tensor $f^{abc}$, but
also admit even-parity ones: $\delta^{ab}$ and $d^{abc}$.
Fundamental generators in real or pseudo-real groups are antisymmetric under exchange,
\begin{equation}
    (T^a)_{i j} \;\xrightarrow{i\leftrightarrow j}\; -(T^a)_{j i} \,
\end{equation}
allowing for a definition of odd-parity color-blocks for appropriate fundamental
representations.  This property is relevant for determining the net
parity under scalar or Majorana fermion exchange.

In what follows we will assume that any vector that appears will be in a
representation that is either in the adjoint or is flavorless. Any scalar that
appears will be dressed in any representation or be flavorless.  Fermions can be
in either the adjoint or fundamental representations. For fermions in the
fundamental of complex representations, the bar will always appear on the
anti-fundamental index and parity will not be defined for exchange for such
color weights.

\subsection{Leg Godt: Color factor definite-parity blocks}
We begin by enumerating the color blocks of definite parity, \ie those for adjoint
or real/pseudo-real fundamental representations.
We define a shorthand for the adjoint color weights on the $s,t$ and $u$ channels as follows:
\begin{align}
    \label{eq:channel-ordering-conventions}
    c^{ff} _s &= f^{a_1 a_2 b}f^{b a_3 a_4},&
    c^{ff} _t &= f^{a_4 a_1 b}f^{b a_2 a_3},&
    c^{ff} _u &= f^{a_3 a_1 b}f^{b a_4 a_2}.
\end{align}
In this convention, the Jacobi identity is given by
\begin{align}
    c^{ff} _s = c^{ff} _t + c^{ff} _u.
\end{align}
In addition to the usual antisymmetric structure constants, the symmetric
structure constants $\delta^{a b}$ and $d^{abc}$ naturally occur in effective
field theories, for instance to capture loop-induced corrections.
Thus, for adjoint-charged particles we include color factors such as $f^{a_1 a_2 b}d^{b a_3 a_4}$ and
$d^{a_1 a_2 b}d^{b a_3 a_4}$.  We define the channel-labeled color factors
dressed with $df,fd$ and $dd$ in a similar manner to
\cref{eq:channel-ordering-conventions}.
Finally, we also consider color factors consisting of fundamental
generators $T^a$ paired with themselves, $d^{abc}$, or $f^{abc}$. 

With these conventions, the color factors with definite parity are given by
\begin{itemize}
 \item[\(( + | +)\)]  Even-Even:  
    \begin{align}
    \label{eq:pp-base-color}
  \mathcal{C}_0 ^{+|+} = \{1,\; c^{Td} _t +c^{Td}_u -c^{dT}_t + c^{dT}_u\}
  \!\!\bigcup_{A,B \in \{f,T\}}\!\! \{c^{AB} _t - c^{AB} _u \} 
  \!\!\bigcup_{A \in \{dd,\delta\delta\}}\!\! \{c^{A} _s, c^{A} _t + c^{A}_u \} .
    \end{align}

  \item[\((- | -)\)]  Odd-Odd:  
    \begin{align}
    \label{eq:mm-base-color}
   \mathcal{C}_0 ^{-|-}=&\bigcup_{A,B \in \{f,T\}} \{c^{AB} _s \} \bigcup_{A,B \in \{f,T\},AB\neq ff} \{ c^{AB} _t + c^{AB} _u \} 
  \nn
  &\bigcup_{A \in \{dd,\delta\delta\}} \{ c^{A} _t - c^{A}_u \}  \bigcup_{B \in \{f,T\}} \{ c^{dB} _t + c^{dB} _u-c^{Bd} _t +c^{Bd}_u \}   .
    \end{align}

 \item[\(( - | + )\)] Odd-Even:  
   \begin{align}
   \label{eq:mp-base-color}
      \mathcal{C}_0 ^{-|+} = \{ c^{fd} _s ,c^{Td} _s ,c^{dT} _t - c^{dT} _u+c^{Td} _t +c^{Td}_u \}.
   \end{align}

 \item[\((+| -)\)] Even-Odd:  
   \begin{align}
   \label{eq:pm-base-color}
      \mathcal{C}_0 ^{+|-} = \{c^{df} _s,c^{dT} _s , c^{dT} _t + c^{dT} _u+c^{Td} _t -c^{Td}_u \} .
   \end{align}
\end{itemize}
In addition to these definite-parity color structures, there are also tensors
which do not have well behaved transformations in either or both pairs.  These
are needed when particles are charged in the fundamental of a complex group, 
or when particle 3 is not charged but particle 4 is -- for instance when 3 is
a photon but 4 is a gluon.  Grouped according to their surviving
exchange properties, they are
\begin{itemize}
 \item[\(( \NA | +)\)]  Undef-Even:  
    \begin{equation}
   \cnp{0} = \{ (T^{b})_{1}{}^{\bar{2}} d^{b a_3 a_4 }, \;
   \delta_{1}{}^{\bar{2}} \delta^{a_3 a_4}  \} \label{eq:color-nap}\, .
    \end{equation}
    
  \item[\(( \NA | -)\)]  Undef-Odd:  
    \begin{equation}
   \cnm{0} =  (T^{b})_{1}{}^{\bar{2}} f^{b a_3 a_4 }  \label{eq:color-nam}  \, .
    \end{equation}

 \item[\((+ | \NA )\)] Even-Undef:  
   \begin{equation}
       \cpn{0} = \{d^{a_1 a_2 b} (T^{b})_{3}{}^{ \bar{4}},\;
       \delta^{a_1 a_2} \delta_{3}{}^{\bar{4}}\} \label{eq:color-pna}
   \end{equation}

 \item[\((- | \NA )\)] Odd-Undef:  
   \begin{equation}
       \cmn{0} = \{f^{a_1 a_2 b} (T^{b})_{3}{}^{ \bar{4}},\; (T^{a_4})_{12} \}
       \label{eq:color-mna}
   \end{equation}

  \item[\((\NA | \NA)\)] Undef:  
\begin{equation}
    \cnn{0} = \{
        (T^{b})_{1}{}^{\bar{2}}
    (T^{b})_{3}{}^{\bar{4}},(T^{a_4})_{1}{}^{\bar{2}},
    \delta_{1}{}^{\bar{2}} \delta_{3}{}^{\bar{4}}
\}
        \label{eq:color-nana}
\end{equation}
\end{itemize}

For ease of notation below, we introduce the fully-general color structures
with possibly-definite parity as
\begin{align}
    \chhfull{h_1}{h_2}{0} = \bigoplus_{q_1 \in \{h_1,\text{NA}\},q_2 \in \{h_2,\text{NA}\}}
    \chh{q_1}{q_2}{0}\,.
    \label{eq:gen-color}
\end{align}
Note that this is purely a formal shorthand, as the particles should always be in
definite representations, thus restricting to specific subsets of the
$\chh{q_1}{q_2}{0}$.

\subsection{Mixing color and kinematics}

Modifications of color building blocks with scalar kinematics naturally show up
when studying the space of effective field theories. It is worth noting
explicitly that they factorize into sums of our building blocks as we discuss
here.

As a key example, one can
capture the space of all higher derivative corrections to maximally
supersymmetric Yang-Mills by just considering all possible scalar modifications
to the color weights, leaving the vector numerators untouched
\cite{Carrasco:2019yyn}:
\begin{align}
    \mathcal{A}_4 ^{\text{sYM + HD}} = \frac{c_s ^{\text{HD}} n_s ^{\text{vec}}}{s}+ \frac{c_t ^{\text{HD}} n_t ^{\text{vec}}}{t}+ \frac{c_u ^{\text{HD}} n_u ^{\text{vec}}}{u},
\end{align}
where $n^{\text{vec}}$ correspond to the vector numerators in $\mathcal{A}_4
^{\text{sYM}}$ and $c^{\text{HD}}$ are color factors modified by scalar
Mandelstams.

We are thus motivated to explore the space of scalar-modified color blocks.
In particular, we are interested in modifications which still result in
definite-parity blocks under the exchange of $1 \leftrightarrow 2$ and $3
\leftrightarrow 4$, which we call $\chh{h_1}{h_2}{D}$.  It is straightforward to
see (either semi-exhaustively via direct computations, or using tools from
classical invariant theory \cite{sturmfels2008algorithms}) that the space
actually cleanly factorizes
\begin{align}
    \chh{h_1}{h_2}{D}= \bigoplus_{q_1,q_2 \in \{-1,1\}} \chh{q_1}{q_2}{0}
    \phh{h_1 q_1}{h_2 q_2}{D},
    \label{eq:dim-color} 
\end{align}
where $\mathcal{C}_0$ correspond to the definite-parity color weights of mass dimension 0,
given in
\cref{eq:pp-base-color,eq:mm-base-color,eq:mp-base-color,eq:pm-base-color}, and
$\phh{h_1}{h_2}{D}$ are the scalar blocks of definite parity with mass
dimension $D$, defined in
\cref{eq:slego-pp-4,eq:slego-mp,eq:slego-pm,eq:slego-mm}.
The factorization trivially extends to NA-type color factors, 
\begin{align}
    \chhfull{h_1}{h_2}{D}= \bigoplus_{q_1,q_2 \in \{-1,1\}} \chhfull{q_1}{q_2}{0}
    \phh{h_1 q_1}{h_2 q_2}{D}.
    \label{eq:dim-color-gen} 
\end{align}
as the sign sums allow the NA terms to appear with arbitrary kinematic dressings.

As a simple example, we show that we span the permutation-invariant color-scalar
mixture
\begin{equation}
    \left( c^{ff}_s t+ c^{ff}_t s\right) ,
\end{equation}
which appears in higher-derivative corrections to bi-adjoint scalar theory
\cite{Bjerrum-Bohr:2012kaa,Cachazo:2013hca,Cachazo:2013iea,Carrasco:2016ldy,Mafra:2016mcc,Carrasco:2016ygv} and Yang--Mills.
As a permutation invariant, this has $(+|+)$ parity under $1 \leftrightarrow 2$ and $3
\leftrightarrow 4$. In terms of our color blocks, this can be written as 
\begin{align}
\label{fourPointAdjointAntisymmetricColorWeight}
\left( c^{ff}_s t+ c^{ff}_t s\right)= \frac{1}{2} \left[ \underbrace{ \left(
c^{ff}_t - c^{ff}_u\right)}_{{\cpp{0}}} \underbrace{s}_{\ppp{2}} +
\underbrace{c^{ff}_s}_{{\cmm{0}}} \underbrace{ \left( t-u \right)}_{\pmm{2}}
\right]\,.
\end{align}

\section{Spin building blocks (Spacetime Parity Conserving)}
\label{sec:spinors}

The strategy is to identify the finite basis of fundamental, $D$-dimensional (parity-even) spinor building blocks which, once
dressed by compatible scalar and  color blocks, span all on-shell four-point
contact amplitudes at arbitrary mass dimension. The reason this works is that, for a given particle content and
little-group weight, there are only finitely many  inequivalent ways of inserting momenta, polarization data, and
Clifford algebra elements into spinor bilinears  before on-shell identities (Dirac equation, momentum conservation,
transversality, and gamma-matrix algebra)  force reduction to a smaller set as we verify explicitly 
through exhaustive on-shell reduction. Any further increase in mass dimension
can only occur through multiplication  by scalar kinematic invariants, which are completely captured by the scalar
polynomial blocks $\mathcal{P}_D$.

The only subtle case is that of four fermions, where one could imagine generating an infinite tower of independent
structures by inserting higher antisymmetric products $\gamma^{\mu_1\cdots\mu_n}$. However, in arbitrary but fixed dimension these are themselves finite up to algebraic equivalence, and in practice we show that it suffices to cap the number of gamma-matrix insertions, as discussed in \cref{sec:merge-fermions}.  Beyond this cap, additional structures are either redundant or reducible to lower ones multiplied by scalar invariants.

The problem of construction and verification of a spanning basis reduces to one of classification.
We now tackle the problem of classifying the space of all possible
\emph{on-shell} spinor blocks -- expressions that involve external
spin-representation data like fermion spinors or vector polarizations -- up to
the overall factors of scalar Mandelstams. Since we are carefully tracking how
different objects transform under exchanging
particles, we will adopt the \emph{Majorana flip condition}
\begin{align}
    \pbarpexpr{ \gamma^{\mu_1 \dots \mu_r} } = t_r\
    \pbarpgenexpr{2}{1}{\gamma^{\mu_1 \dots \mu_r}}\,,
    \label{eq:maj-flip}
\end{align}
with $t_r = \pm 1$, $t_2 = - t_0$, $t_3 = - t_1$ and $t_{r+4} = t_{r}$, as the
definition for how spinor bilinears change under exchanging particles. The signs
$t_r$ depend on one's choice of charge-conjugation matrix and gamma-matrix
conventions. We classify the parity of our spinors blocks with the $D=4$ choice:
$t_0=-1$ and $t_1 = +1$. Our conventions agree with the standard SUSY amplitudes
literature, and require no additional bookkeeping when projecting to spinor
helicity variables (\eg both sides of $\pbarpgenexpr{i}{j}{} \leftrightarrow
\left<ij\right>$ are antisymmetric under Majorana exchange of $i$ and $j$). We emphasize that
we make this choice of $D=4$ and  $t_0, t_1$ solely for the purposes of
organizing our spinor blocks into families of definite parity; the general
structure of the overall analysis and the basis spinors blocks we obtain remain
valid in all dimensions, up to parity conventions. For construction 
considerations in general $D$ one simply carries along the undetermined
$\{t_r\}$.

Generically, the fermionic spinor building blocks are built from elements of the
$D$-dimensional Clifford algebra sandwiched between two spinors.  The even
$D$-dimensional Clifford algebra is spanned by the antisymmetric product of up
to $D$ $\gamma$ matrices:
\begin{align}
    \Gamma_{\text{even}}^A = \{ 1, \gamma^{\mu}, \gamma^{\mu \nu}, \dots, \gamma^{\mu_1 \dots
    \mu_D}\}
    \label{eq:clifford-alg}
\end{align}
with
\begin{align}
    \gamma^{\mu \nu} = \frac{1}{2} \left(\gamma^\mu \gamma^\nu - \gamma^\nu
    \gamma^\mu\right) \qquad \gamma^{\mu_1 \dots \mu_n} = \frac{1}{n!} \left(
        \gamma^{\mu_1} \dots \gamma^{\mu_n} - \gamma^{\mu_2} \gamma^{\mu_1}
        \dots \gamma^{\mu_n} + \dots \right)\,,
        \label{eq:anti-ex}
\end{align}
while the odd $D$-dimensional basis is only ``half-sized'' due to duality relations
\begin{align}
    \Gamma_{\text{odd}}^A = \{ 1, \gamma^{\mu}, \gamma^{\mu \nu}, \dots, \gamma^{\mu_1 \dots
    \mu_{(D-1)/2}}\}\,.
    \label{eq:clifford-alg-odd}
\end{align}
Because we are cataloging properties in general dimension, we will simply refer to the
Clifford algebra basis as $\Gamma^A$, and will not be exploiting any particular
properties of even or odd dimensions.
Thus, the most general fermionic spin building blocks are
$\pbarpgenexpr{i}{j}{\Gamma^A}$.

\subsection{Leg Godt: Spinor multilinear definite-parity blocks}

We will organize our spin blocks into four families, distinguished by the natures of legs $3$ and $4$:
\begin{enumerate}
\item Two fermions + two scalars (\cref{sec:twoFtwoSc});
\item Four fermions (\cref{sec:fourFermions});
\item Two fermions + two vectors (\cref{sec:twoFtwoV});
\item Two fermions + one scalar + one vector (\cref{sec:TwoFVS}).
\end{enumerate}
In each case, we organize our spinor blocks into families that have definite parity under
the Majorana exchange $1 \leftrightarrow 2$ (and independently $3
\leftrightarrow 4$ where relevant). We label our spinor blocks as 
\begin{align}
    \spinBlock{\text{species, \# $\gamma$s, mass dim, other}}{\text{sign}(1 \leftrightarrow 2)}{\text{sign}(3 \leftrightarrow 4)}
\end{align}
where we use the operator engineering dimensions in 4D as the mass dimension
counting, \ie we take the mass dimension of a spinor to be $3/2$ and that of
momenta and polarization vectors to be $1$.

In the following subsections, we classify the space of all unique spinor blocks
consistent with their corresponding kinematic interactions, up to overall factors
of scalar Mandelstams. We start by building all possible Lorentz invariant
spinor bilinears that potentially contain momenta or polarization vectors 
at a given dimension. We then impose on-shell conditions, momentum conservation,
and the Dirac equation to prune linearly dependent terms. For the case of vectors,
we also impose transversality and gauge invariance as additional constraints. Each of
the remaining functions forms a valid on-shell spinor block. We then remove
degeneracies that arise from lower-mass-dimension basis elements multiplied by
functions of scalar Mandelstams $\mathcal{P}_{D}$ to obtain a minimal basis of our
spinor blocks. For each particle content, the exhaustiveness of the resulting bases is 
established by this explicit construction, with additional high-dimensional checks 
discussed on a case-by-case basis below.

\subsection{Two fermions + two scalars (2F+2S)}
\label{sec:twoFtwoSc}
Naively, the most general spinor bilinear we can construct for interactions involving two
fermions and two scalars is given by
\begin{align}\label{eq:2F2SBrute}
    \pbarpexpr{ \slashed{k}_{i_1} \dots \slashed{k}_{i_n} },
\end{align}
where $i_{j} \in \{1,2,3,4\}$.  However, we can always reduce such a term by:
\begin{enumerate}
    \item Removing all occurrences of $k_4$ via momentum conservation.
    \item Using $\gamma$ anti-commutation $\{\gamma^\mu, \gamma^\nu\} = 2
        \eta^{\mu \nu}$ to move all $\slashed{k}_2$ to the rightmost end of the
        contraction, and then the Dirac equation to turn it into a mass.
        Similarly, all $\slashed{k}_1$ can be moved left and then removed from
        the $\gamma$ contractions.
    \item Reducing the remaining chain involving only $\slashed{k}_3$ via
        $\slashed{k}_3 \slashed{k}_3 \to m_3^2$.
\end{enumerate}
Thus the minimal spinor bilinears in this case are 
\begin{align}
    \pbarpexpr{ } \hspace{1cm} \text{and} \hspace{1cm}  \pbarpexpr{ \slashed{k}_3 }.
\end{align}
We can make their parity more apparent by suggestively writing them as 
\begin{itemize}
    \item[$(+|-)$] Even-Odd:
        \begin{align}
            \spinBlockS{6}{\tone}{-} &= \frac{1}{2}(\pbarpk{3} - \pbarpk{4})
            \label{eq:spin-k-m} \,, 
        \end{align}

            \item[$(-|+)$] Odd-Even:
        \begin{align}
            \spinBlockS{5}{\tzero}{+} &= \pbarp{} 
            \label{eq:spin-empty} \, .
        \end{align}
\end{itemize}
Because the mass dimension sufficiently identifies the two terms, we omit
the $\gamma$-count subscript on these blocks.

\subsection{Four fermions (4F)}
\label{sec:fourFermions}
The most general spinor bilinear we can construct for interactions involving
four fermions is given by
\begin{align}\label{eq:4FBrute}
   \left( \pbarpexpr{  \slashed{p}_1 \dots \slashed{p}_m \Gamma^{A} }
   \right) \left( \pbarpgenexpr{3}{4}{  \slashed{q}_1 \dots \slashed{q}_n \Gamma_{A} }\right)\,,
\end{align}
with $p$s and $q$s drawn from $\{k_1,k_2,k_3,k_4\}$.
We can mimic our above analysis of 2F+2S and reduce the ansatz such that we
always have $m,n \leq 1$. As such, the four-fermion spinor basis is covered by
\begin{subequations}
\begin{align}
    \spinBlockF{A,6}{t_A}{t_A} &= \pbarpexpr{ \Gamma^A }\ \pbarpgenexpr{3}{4}{ \Gamma_A }
    \label{eq-app:4spin-no-k}\\
    \spinBlockF{A,7,r}{t_{A+1}}{-t_A} &=\pbarpexpr{ (\slashed{k}_3-\slashed{k}_4) \Gamma^A }\ \pbarpgenexpr{3}{4}{ \Gamma_A }
    \label{eq-app:4spin-k3}\\
    \spinBlockF{A,7,l}{-t_A}{t_{A+1}} &=\pbarpexpr{ \Gamma^A }\
    \pbarpgenexpr{3}{4}{ (\slashed{k}_1-\slashed{k}_2) \Gamma_A }
    \label{eq-app:4spin-k1} \\
    \spinBlockF{A,8}{-t_{A+1}}{-t_{A+1}} &=\pbarpexpr{ (\slashed{k}_3-\slashed{k}_4) \Gamma^A }\ \pbarpgenexpr{3}{4}{ (\slashed{k}_1-\slashed{k}_2) \Gamma_A }
    \label{eq-app:4spin-kk}\,,
\end{align}
\label{eq-app:4spin-basis}
\end{subequations}
where by $\Gamma^A \cdots \Gamma_A$ we mean any pairing of an element from
\cref{eq:clifford-alg} (or \cref{eq:clifford-alg-odd}) with itself. Here, the
signatures for $t_A$ in the expressions above result from incorporating the
exchange of fermion momentum into our definition of particle exchange, and
depend on the gamma matrix conventions, as mentioned at the beginning of this
section. 

We note that the basis above spans all the spin blocks in \emph{arbitrary}
spacetime dimensions. However, in specific dimensions it is often possible to
relate basis elements to each other using Fierz identities.  In particular, in
4D, Fierz identities allow us to express all of our basis elements listed above
in terms of ones with a summation over one gamma matrix at most. We provide a
detailed discussion of the basis elements in 4D in \cref{sec:4F4D}.

\subsection{Fermion pair and vector pair (2F+2V)}
\label{sec:twoFtwoV}
Next, we give a basis of spinor bilinears that span all
gauge-invariant combinations encoding the coupling of two vectors to two
fermions. We restrict ourselves to (not necessarily identical) massive fermions
and massless vectors in this work, although the formalism easily generalizes to
massive vectors by including a longitudinal spurion and Stueckelberg-type
blocks. 

The spinor blocks here are composed of spinor bilinears along with
Lorentz products involving polarization vectors. The primary property we
require of these spinor blocks is gauge invariance of the
external vectors, which can be phrased as the constraint
\begin{equation}
    n \big|_{\pol_i \to k_i} = 0,
    \label{eq:gauge-inv}
\end{equation}
where $i$ can be either of the two external vectors. It is straightforward to
find valid building blocks using a brute-force ansatz. We can exploit on-shell
kinematics, just as we did for the 2F+2S and 4F case, to reduce any spinor
bilinear $ \pbarpexpr{ \slashed{v}_1 \dots \slashed{v}_n }$ in terms
of one with at most three $\gamma$ matrices. The possibility of two additional
$\gamma$ matrices is due to the potential presence of $\slashed{\epsilon}$
inside the spinor bilinears. Hence, any spinor block corresponding to two
fermions and two vectors can be written in terms of products of spinor bilinears 
\begin{align}
\label{eq:set of spinor bilinears}
   \{ &\pbarpexpr{ },\; \pbarpexpr{ \slashed{k}_3 },\; \pbarpexpr{ \slashed{\epsilon}_3}, 
   \pbarpexpr{ \slashed{\epsilon}_4},\; 
   \pbarpexpr{ \slashed{\epsilon}_3 \slashed{k}_3},\; 
\pbarpexpr{ \slashed{\epsilon}_4 \slashed{k}_3},\;
\pbarpexpr{ \slashed{\epsilon}_3 \slashed{\epsilon}_4},\;
\pbarpexpr{ \slashed{\epsilon}_3 \slashed{\epsilon}_4 \slashed{k}_3 } \}
\end{align}
The most general ansatz we can write at mass dimension $d$ is given by
\begin{align}
    \vfBlock{d}^{\text{ansatz}}= &\sum_{v_i  \in \mathcal{P}_{d-5}} a_{1i}v_i (\pol_3 \cdot \pol_4) \pbarp
    + \sum_{i,j \in \{1,2\}, v_k \in \mathcal{P}_{d-7} } a_{2ijk}v_k (\pol_3 \cdot k_i)(\pol_4 \cdot k_j)
    \pbarp \nn
    &+\sum_{v_i  \in \mathcal{P}_{d-6}} b_{1i} v_i (\pol_3 \cdot \pol_4) \pbarpexpr{ \slashed{k}_3 }
    + \sum_{i,j \in \{1,2\}, v_k \in \mathcal{P}_{d-8} } b_{2ijk}v_k (\pol_3 \cdot k_i)(\pol_4 \cdot k_j) \pbarpexpr{ \slashed{k}_3 } \nn
    &+\sum_{v_i  \in \mathcal{P}_{d-6}, j \in \{1,2\}} b_{3ij} v_i  (\epsilon_4 \cdot k_j)\pbarpexpr{ \slashed{\epsilon}_3 } +\sum_{v_i  \in \mathcal{P}_{d-6}, j \in \{1,2\}} b_{4ij} v_i  (\epsilon_3 \cdot k_j)\pbarpexpr{ \slashed{\epsilon}_4 }, \nn
&+\sum_{ j \in \{1,2\}, v_i \in \mathcal{P}_{d-7}} c_{1ij} v_i  (\epsilon_4 \cdot k_j)\pbarpexpr{ \slashed{\epsilon}_3 \slashed{k}_3 }+\sum_{ j \in \{1,2\}, v_i \in \mathcal{P}_{d-7}} c_{2ij} v_i  (\epsilon_3 \cdot k_j)\pbarpexpr{ \slashed{\epsilon}_4 \slashed{k}_3 } \nn
    & +\sum_{v_i \in \mathcal{P}_{d-5}} c_{3i} v_i \pbarpexpr{ \slashed{\epsilon}_3 \slashed{\epsilon}_4 }  +\sum_{v_i \in \mathcal{P}_{d-6}} c_{4i} v_i \pbarpexpr{ \slashed{\epsilon}_3 \slashed{\epsilon}_4 \slashed{k}_3 }.
    \label{eq:vec-ans}
\end{align}
Here, we skip over the summations involving $\mathcal{P}_{m}$ with $m <0$. The
summations involving $(\pol_{3/4} \cdot k_i)$ (like in the second terms) are
only over $\{1,2\}$ due to a combination of momentum conservation and transversality.

Alternatively, a natural way to construct gauge-invariant spinor bilinears is to
build them out of the linearized field strength:
\begin{equation}
\label{eq:linearizedF}
    F_i^{\mu \nu} \equiv k_i^\mu \pol_i^\nu - k_i^\nu \pol_i^\mu\,.
\end{equation}
which trivially satisfies \cref{eq:gauge-inv}.  It turns out that up through at
least mass dimension 23, all gauge invariant solutions of \cref{eq:vec-ans} are
easily writable in terms of $F_3$ and $F_4$.  This is well beyond 
where novel contractions of field strengths stop being possible (at mass
dimension 11), so is very strong evidence that all possible solutions are
writable in terms of field strengths.

We then turn to analyzing the solutions and finding a minimal basis.
The analysis begins at dimensions $5$ and $6$, where imposing gauge invariance
leaves us with no non-trivial solutions . The lack of a possible 2V 2F operator
for these mass dimensions is an often discussed result \cite{SIMMONS1990471}.
However, from the linearized-field-strength perspective this isn't surprising, as
the simplest objects we can construct from $\pbarpexpr{, }, F_3$ and
$F_4$ only occur at dimension 7.

Non-trivial gauge-invariant solutions exist starting at mass dimension $7$ where
there are $3$.  We further find $11$ solutions at dimension $8$ and $27$
solutions at dimension $9$. However, $6$ of the dimension-8 solutions are simply
the fermion masses multiplying the dimension-7 solutions.  Thus there are only 5
novel tensor structures at dimension 8.  Similarly, at dimension 9 all but 2 of
the solutions can be written in terms of lower-mass-dimension tensor structures
multiplied by $\mathcal{P}_{d}$. We express our basis elements in terms of the
linearized field strength, $F^{\mu \nu}$, defined in \cref{eq:linearizedF}. We
normalize $\slashed{F}_i$ to be
\begin{align}
    \slashed{F} \equiv \frac{1}{4} \gamma^{\mu \nu} F_{\mu \nu} = \slashed{k} \slashed{\epsilon}.
\end{align}
The solutions can be divided into families of definite parity as:
\begin{itemize}
    \item[$(+|+)$] Even-Even: None
    \item[$(-|+)$] Odd-Even:
        \begin{align}
            \spinBlockV{0,7}{-}{+} &= \tr(F_3,F_4) \pbarp{} \,,\\
            \spinBlockV{0,9}{-}{+} &= (k_1 - k_2) \cdot F_3 \cdot F_4 \cdot (k_1 - k_2) \pbarp{}\,,\\
            \spinBlockV{1,8}{-}{ +} &= (k_1 - k_2)^{\rho} \left(F_{4,\rho \mu} \pbarpf{3} + F_{3,\rho \mu} \pbarpf{4} \right)\,, \\   
            \spinBlockV{3,8}{-}{+} &= \pbarpTrip F_3{}_{\mu}{}^{\sigma}
            F_4{}_{\sigma \nu} (k_3 - k_4)_{\rho}\,,\\
        \spinBlockV{4,7}{\tzero}{+} &= \pbarpexpr{ \left( \slashed{F}_{3} \slashed{F}_{4} + \slashed{F}_{4} \slashed{F}_{3} \right) }  
         .
        \label{eq:fslashfslash}
        \end{align}

    \item[$(+|-)$] Even-Odd:
        \begin{align}
            \spinBlockV{1,8}{+}{-} &=  \tr(F_3,F_4) \left(\pbarpk{3} - \pbarpk{4}\right) \,,\\
          \spinBlockV{2,7}{+}{-} &=\pbarpDoub F_3{}_{\mu}{}^{\rho}F_4{}_{\rho \nu}\,,\\
         \spinBlockV{3,8}{+}{-} &= \pbarpTrip\ F_3{}_{\mu \nu} F_4{}_{\rho \sigma} (k_1 - k_2)^{\sigma} - (3 \leftrightarrow 4) \,.  
        \end{align}

    \item[$(-|-)$] Odd-Odd:  
        \begin{align}
            \spinBlockV{1,8}{-}{-} &= (k_1 - k_2)^{\rho} \left(F_{4,\rho \mu} \pbarpf{3} - F_{3,\rho \mu} \pbarpf{4}\right) \,, \\
            \spinBlockV{2,9}{-}{-} &= \pbarpexpr{ \slashed{F}_3 } (k_1 - k_2) \cdot F_4 \cdot k_3 - (3 \leftrightarrow 4)\,.
       \end{align}
\end{itemize}
Notably, $\spinBlockV{4,7}{\tzero}{+}$ (\cref{eq:fslashfslash}) appears to violate the bilinear
restrictions set out in \cref{eq:set of spinor bilinears}.  However, applying
momentum conservation on the $\slashed{k}_4$ within $\slashed{F}_4$ allows the
fourth $\gamma$ to be removed
\begin{equation}
    \spinBlockV{4,7}{\tzero}{+} \to (m_1 + m_2) \pbarpTrip \pol_{3{\mu}}
    \pol_{4{\nu}} k_{3{\rho}}+ \dots
\end{equation}
meaning it does in fact stem from \cref{eq:set of spinor bilinears}.

We have two pieces of evidence that the above basis is complete.  First, we have
checked that \emph{all possible} contractions of $F_3$ and $F_4$ into $k$s,
$\gamma$s, and each other are covered by our basis, possibly multiplied by scalar
kinematic functions.  For instance, one of the highest-mass-dimension
contractions of $F$s is
\begin{align}
    \left(k_3 \cdot F_4 \cdot (k_1 - k_2)\right) 
    \left(k_4 \cdot F_3 \cdot (k_1 - k_2)\right) \pbarp{}
    &= \frac{1}{8}\left((m_1^2 - m_2^2)^2 - (t - u)^2\right)
    \spinBlockV{0,7}{-}{+} \notag \\
    &\quad - \frac{1}{2} s\, \spinBlockV{0,9}{-}{+}\,.
\end{align}
Second, we have explicitly constructed \cref{eq:vec-ans} for $d \le 23$, using finite field methods \cite{Mangan:2023eeb}, and
found that all gauge-invariant solutions are spanned by $\phh{q_1}{q_2}{d_1}
\otimes \spinBlockV{i,d_2}{p_1}{p_2}$.

\subsection{Fermion pair, one scalar, and one vector (2F+1S+1V)}
\label{sec:TwoFVS}
We finally list all the gauge-invariant spinor blocks corresponding to interactions of two fermions, a scalar and one vector. There is nothing new conceptually when classifying the spinor blocks for this case, so we skip the analysis and directly list them below:
\begin{align}
    \spinBlockSV{0,8}{+} &= (k_1 - k_2) \cdot F_4 \cdot k_3 \pbarp{}\\
    \spinBlockSV{1,7}{-} &= (k_1 - k_2)_\mu \pbarpf{4}\\
    \spinBlockSV{1,7}{+} &= k_{3\mu} \pbarpf{4}\\
    \spinBlockSV{2,6}{+} &= \pbarpexpr{ \slashed{F}_4 }\,.
\end{align}
Here, we take the scalar and vector to be particles $3$ and $4$, respectively.
We note that the exchange signatures listed above correspond to the exchange of
fermions.
Note that a scalar \emph{field} has operator-dimension 1, but without a
corresponding derivative does not contribute anything to the contact amplitude.

\section{Merging to full amplitudes}
\label{sec:merge}
We now proceed with the assembly of our modular building blocks. 
We showed that we can organize the spaces of each of these blocks into families
of definite exchange parity. Moreover, we demonstrated that even when considering
contributions that look like mixing color and kinematic weights, the building
blocks fully factorize.  Therefore, to assemble any $D$-dimension amplitude, we
simply take
\begin{align}\label{eq:brute merge}
    \mathcal{A}_D ^{(h_1 | h_2)} (1234) \in 
    \bigoplus_{\substack{d_1  + d_2=D\\ p_i ,q_i \in \{-1,1\} }} \phh{p_1 q_1 h_1}{p_2 q_2
h_2}{d_1} \chhfull{p_1}{p_2}{0}  \spinBlock{\dots,d_2}{q_1}{q_2},
\end{align}
with the remaining consideration that $h_1$ and $h_2$ are chosen to imbue the
amplitude with the required particle statistics, \ie Fermi symmetry under
exchange of identical fermions, and Bose symmetry under exchange of identical
scalars/vectors.  We continue to use the $\bigoplus$ to remind that the RHS of
\cref{eq:brute merge} is a basis of objects that require specific choices of
Wilson coefficients to match any particular amplitude. We note that certain elements of the direct sum may be forbidden on account of violation of internal symmetries of the theory, such as charge conservation.

\subsection{Two fermions and two scalar}
\label{sec:merge-scalars}
Consider a $D$-dimensional amplitude corresponding to two fermions and two scalars. We will take our representative particles to be $1,2$ as fermions and $3,4$ as scalars. We need this amplitude to have $-$ parity under $1 \leftrightarrow 2$ and $+$ parity under $3 \leftrightarrow 4$.
As such, our amplitude will be an element of 
\begin{align}\label{eq:scalar merge}
    \mathcal{A}_D ^{(- |+)} (\psi^a \psi^b \phi^c \phi^d) &\in \bigoplus
    _{\substack{d_1  + d_2=D\\p_i,q_i \in \{ -1 ,1 \}}}
\phh{-p_1 q_1}{p_2 q_2}{d_1} \chhfull{p_1}{p_2}{0} \spinBlockS{0,d_2}{q_1}{q_2} \nn
    &=\bigoplus_{p_1, p_2 \in \{ -1 ,1 \} }\!\!
    \phh{p_1}{p_2}{D-5} 
    \chhfull{p_1}{p_2}{0} 
    \spinBlockS{0,5}{-}{+} \nn
    &\quad \bigoplus_{p_1, p_2 \in \{ -1 ,1 \} }
    \phh{-p_1}{-p_2}{D-6}
    \chhfull{p_1}{p_2}{0}
    \spinBlockS{0,6}{+}{-} ,
\end{align}
where $\chhfull{p_1}{p_2}{0}$ is defined in \cref{eq:gen-color}.  Note that the
case of fermions not being exchangeable is covered by the $\NA$ components
included in $\chhfull{p_1}{p_2}{0}$.  The remaining sign sums come together to
cover all possible kinematic dressings.

\subsection{Four fermions}
\label{sec:merge-fermions}
Next, we consider $D$-dimensional four-fermion amplitudes. In
particular, we deal with the case where fermions $1$ and $2$ belong to 
one family, and fermions $3$ and $4$ belong to a possibly-different family. We need such a
four-fermion amplitude to have $-$ parity under $1 \leftrightarrow 2$ and $-$
parity under $3 \leftrightarrow 4$. In this case, \cref{eq:brute merge} reduces
to
\begin{align}\label{eq:fermion merge}
    \mathcal{A}_D ^{(- |- )} (\psi^a \psi^b \psi^c \psi^d) 
    &\in 
    \bigoplus _{\substack{d_1  + d_2=D\\ p_i, q_i \in \{ -1 ,1 \}}}
    \phh{-p_1 q_1}{-p_2 q_2}{d_1}
    \chhfull{p_1}{p_2}{0}
    \spinBlockF{A,d_2}{q_1}{q_2} \nn
    &=\bigoplus _{p_i \in \{ -1 ,1 \}}
    \phh{-t_A p_1}{-t_A p_2}{D-6}
    \chhfull{p_1}{p_2}{0}
    \spinBlockF{A,6}{t_A}{t_A} \nn
     &\quad \bigoplus _{p_i \in \{ -1 ,1 \}}
    \phh{-p_1 t_{A+1}}{ p_2 t_A}{D-7} 
    \chhfull{p_1 }{ p_2}{0} 
    \spinBlockF{A,7,r}{t_{A+1}}{-t_A} \nn
     &\quad \bigoplus _{p_i \in \{ -1 ,1 \}}
     \phh{p_1 t_A}{ -p_2 t_{A+1}}{D-7} \chhfull{p_1 }{ p_2}{0}
     \spinBlockF{A,7,l}{-t_A}{t_{A+1}} \nn
     &\quad \bigoplus _{p_i \in \{ -1 ,1 \}}
     \phh{p_1 t_{A+1}}{ p_2t_{A+1}}{D-8} \chhfull{p_1  }{ p_2}{0}  \spinBlockF{A,8}{-t_{A+1}}{-t_{A+1}}
\end{align}
where $\chhfull{p_1}{p_2}{0}$ is defined in \cref{eq:gen-color} and $n_{4\psi}$
are the basis of four fermion spinor blocks outlined in \cref{sec:fourFermions}.
Again the potential distinctness of the fermions is implicit through
$\NA$ color factors. 

\subsection{Two fermions and two vectors}
\label{sec:merge-vectors}

Now we come to the case of two massless vectors and a pair of massive
fermions, interacting via a quartic vertex. We take particles $1$ and $2$ to be
fermions and particles $3$ and $4$ to be the massless vectors. We need any
amplitude corresponding to the interactions between these particles to have $-$
parity under $1 \leftrightarrow 2$.
If the vectors are the same species, then they must have $+$ parity under $3
\leftrightarrow 4$, leading to
\begin{align}
    \mathcal{A}_D ^{(- |+)} (\psi^a \psi^b A^c A^d) \in 
    &\bigoplus _{\substack{d_1  + d_2=D\\ p_i ,q_i \in \{-1,1\} }}
    \phh{p_1 q_1 }{ p_2 q_2 }{d_1} 
    \chhfull{p_1 }{ p_2}{0}  
    \spinBlockV{A,d_2}{q_1}{q_2} \nn
    =& \bigoplus _{p_i  \in \{-1,1\} }\phh{p_1 }{ -p_2 }{D-8} \chhfull{p_1 }{ p_2}{0}  \spinBlockV{1,8}{-}{-} 
     \bigoplus _{p_i  \in \{-1,1\} }\phh{p_1 }{ -p_2 }{D-9} \chhfull{p_1 }{ p_2}{0}  \spinBlockV{2,9}{-}{-}  \nn
    & \bigoplus _{p_i  \in \{-1,1\} }\phh{-p_1 }{ -p_2 }{D-8} \chhfull{p_1 }{ p_2}{0}  \spinBlockV{1,8}{+}{-}
    \bigoplus _{p_i  \in \{-1,1\} }\phh{-p_1 }{ -p_2 }{D-7} \chhfull{p_1 }{ p_2}{0}  \spinBlockV{2,7}{+}{-}\nn
    &\bigoplus _{p_i  \in \{-1,1\} }\phh{-p_1 }{ -p_2 }{D-8} \chhfull{p_1 }{ p_2}{0}  \spinBlockV{3,8}{+}{-} 
    \bigoplus _{p_i  \in \{-1,1\} }\phh{p_1 }{ p_2 }{D-7} \chhfull{p_1 }{ p_2}{0}  \spinBlockV{0,7}{-}{+} \nn
    & \bigoplus _{p_i  \in \{-1,1\} }\phh{p_1 }{ p_2 }{D-9} \chhfull{p_1 }{ p_2}{0}  \spinBlockV{0,9}{-}{+}  
    \bigoplus _{p_i  \in \{-1,1\} }\phh{p_1 }{ p_2 }{D-8} \chhfull{p_1 }{ p_2}{0}  \spinBlockV{1,8}{-}{+} \nn
    &  \bigoplus _{p_i  \in \{-1,1\} }\phh{p_1 }{ p_2 }{D-8} \chhfull{p_1 }{ p_2}{0}  \spinBlockV{3,8}{-}{+}  
    \bigoplus _{p_i  \in \{-1,1\} }\phh{p_1 }{ p_2 }{D-7} \chhfull{p_1 }{ p_2}{0}  \spinBlockV{4,7}{-}{+}.
\end{align}
Similar to the fermionic exchange, the case where the two vectors are different
species is implicitly covered by the $\chh{h_1}{\NA}{0}$ color factors within
$\chhfull{h_1}{\pm}{0}$.

\subsection{Fermion pair, one scalar, one vector}
Amplitudes corresponding to two fermions, one scalar and one vector particle only have definite (negative exchange) parity under the exchange of fermions $1 \leftrightarrow 2$, allowing for all other possibilities relevant to a given mass dimension.
\begin{align}\label{eq:merge-scalar-vector}
    \mathcal{A}_D ^{(-)} (\psi^a \psi^b \phi A) \in 
&\bigoplus_{\substack{d_1  + d_2=D\\ p_i ,q,r \in \{-1,1\}}} 
\phh{-p_1 q }{r}{d_1} 
\chhfull{p_1}{p_2}{0}  
\spinBlockSV{A,d_2}{q} \nn
=&\bigoplus_{ p_i ,r \in \{-1,1\} } 
\phh{-p_1 }{r}{D-8} 
\chhfull{p_1}{p_2}{0}  
\spinBlockSV{0,8}{+} 
\bigoplus_{ p_i ,r \in \{-1,1\} } 
\phh{p_1 }{r}{D-7} 
\chhfull{p_1}{p_2}{0}  
\spinBlockSV{1,7}{-}\nn
&\bigoplus_{ p_i ,r \in \{-1,1\} } 
\phh{-p_1}{r}{D-7} 
\chhfull{p_1}{p_2}{0}  
\spinBlockSV{1,7}{+}
\bigoplus_{ p_i ,r \in \{-1,1\} } 
\phh{-p_1 }{r}{D-6} 
\chhfull{p_1}{p_2}{0}  
\spinBlockSV{2,6}{+},
\end{align}
where $n_{sv}$ are the spinor bilinears defined in \cref{sec:TwoFVS}.

\section{Double Copy}
\label{sec:DC}
Over the past few decades, the double copy has emerged as a unifying perspective
which relates predictions in theories that might seem entirely unrelated, such
as gauge theories with a finite number of contact terms, and gravitational
theories with an infinite tower of higher-derivative interactions. While it has
long been known that free graviton polarizations factor into products of gauge
theory polarizations, the surprise was that this factorization extends to full
tree-level amplitudes, graph by graph and to all multiplicities. First
discovered in the context of Yang-Mills amplitudes double-copying into
gravitational theories these ideas now span a wide web of relations—from pure gauge theories to the
bosonic components of the superstring and effective theories such as Born-Infeld
and Dirac–Born–Infeld.

The modularity and manifest gauge invariance of our building blocks makes them
naturally suited to this broader double-copy structure. In particular, the
process of combining kinematic weights with kinematic weights mirrors the above
merging procedure of combining kinematic weights with color weights to form
predictions in gauge theory. This is accomplished by ensuring that the resulting
amplitude transforms correctly under exchange symmetry: antisymmetric for
identical fermions and symmetric for identical bosons.  Under those constraints
our double-copy predictions are spanned by:
\begin{equation}
\boxed{
\text{Double-Copy contact} = [ \text{scalar block}] \times [ \text{spin block}^\text{(A)}] \times [ \text{spin block}^\text{(B)} ] \, . }
\label{genDC}
\end{equation}

It is important to emphasize that in our framework, the main challenge is not
enforcing color-kinematics duality. That structure is already reflected in the
modular decomposition itself. The challenge is interpretive: understanding what
a given product of building blocks corresponds to in the double-copy theory. In
other words, given the freedom to multiply gauge-invariant, little-group
covariant components, we must determine which combinations yield meaningful
gravitational states or interactions.

A useful double-copy should satisfy several criteria:
\begin{enumerate}
    \item It should reduce complex calculations to combinations of simple, universal building blocks.
    \item It should lift linearized gauge invariance to linearized diffeomorphism invariance.
    \item It should preserve factorization on physical channels.
    \item It should respect spin-statistics.
\end{enumerate}
For higher-derivative local operators at four points, our modular framework
largely satisfies these by construction. Gauge invariance (point 2) is built
into the blocks, and spin-statistics (point 4) is enforced through the symmetry
properties of the combinations we allow. Because contact terms need not
factorize on physical poles, point 3 is not a constraint at this stage. Thus,
the real subtlety lies in point 1: choosing the combination that builds the
object you want to predict in the double-copy theory. With that in mind, we
begin by reviewing how states combine under kinematic double-copy, then discuss
how our framework relates to the traditional antisymmetric-adjoint double-copy
of
KLT \cite{Kawai:1985xq}, BCJ \cite{Bern:2008qj,Bern:2010ue,Bern:2019prr} and
CHY \cite{Cachazo:2013iea,Cachazo:2013hca,Edison:2020ehu}.

\subsection{Double-copy of states in $D$ Dimensions}
 
 The kinematic double copy constructs composite gravitational states
 by tensoring on-shell states from two single-copy gauge theories. In any spacetime
 dimension $ D$, each on-shell particle transforms under the little group $
 \mathrm{SO}(D{-}2) $, and the resulting double-copy state space is built from
 tensor products of little-group representations~\cite{Bern:2019prr}.

\noindent
\textbf{Scalars}, being singlets under $ \mathrm{SO}(D{-}2) $, do not affect the
little-group structure of their partners; they merely shift the mass dimension
of the composite state. Thus, scalar $\otimes$ anything leaves the spin
unchanged, and can be used to generate massive or higher-derivative corrections
without modifying spin content.

\noindent
\textbf{Gluons}, or more generally massless vectors, transform in the vector
representation of $ \mathrm{SO}(D{-}2) $. The double copy of two gluons
produces:
\begin{itemize}
    \item A symmetric traceless tensor: corresponding to the \textbf{graviton},
    \item A scalar (from the trace): the \textbf{dilaton},
    \item An antisymmetric tensor: the \textbf{Kalb--Ramond 2-form}.
\end{itemize}
These arise from the decomposition
\begin{equation}
V \otimes V = (\mathrm{Sym}_0^2 V \oplus \text{Trace}) \oplus \Lambda^2 V= \text{graviton} \oplus \text{dilaton} \oplus B_{\mu\nu}\,,
\end{equation}
where $\mathrm{Sym}_0^2 V$ denotes the symmetric traceless rank-2 tensor (the
graviton), the trace part yields a little-group scalar (the dilaton) and the
$\Lambda^2 V$ is the antisymmetric 2-form representation (corresponding to the
Kalb-Ramond field $B_{\mu\nu}$.

\noindent
\textbf{Fermions} transform in spinor representations of $ \mathrm{SO}(D{-}2) $, and tensor in the following characteristic ways:
\begin{itemize}
    \item Scalar $\otimes$ fermion produces a fermion (unchanged spin),
    \item Fermion $\otimes$ vector yields spin-3/2–like states (gravitini),
    \item Fermion $\otimes$ fermion yields bosonic states:
    \begin{itemize}
        \item In even $ D $, same-chirality fermions give antisymmetric products including $p$-forms (e.g., RR fields in Type IIB),
        \item Opposite-chirality fermions give symmetric products, including scalars and vectors (e.g., RR fields in Type IIA).
    \end{itemize}
\end{itemize}
These patterns obey spin-statistics: fermionic outputs require exactly one
fermion in the tensor product; all other combinations yield bosonic states.
The state-level double copy is summarized in \cref{tab:state-table}.
\begin{table}[h]
\begin{center}
    \begin{tabular}{lll}
        \toprule
        \textbf{Left copy} & \textbf{Right copy} & \textbf{Double copy output} \\
        \midrule
        Scalar     & Scalar     & Scalar \\
        Scalar     & Fermion    & Fermion \\
        Vector     & Scalar     & Vector \\
        Vector     & Vector     & Graviton $\oplus$ Dilaton $\oplus$ B-field \\
        Fermion    & Fermion    & Scalar $\oplus$ Vector $\oplus$ Forms (depends on chirality) \\
        Vector     & Fermion    & Gravitino \\
        \bottomrule
    \end{tabular}
\end{center}
\caption{Summary of double copies between various particle states.}
\label{tab:state-table}
\end{table}

The procedure can naturally be extended to construct higher-spin or massive
states by chaining together single-copy representations \footnote{See, e.g.~Ref.~\cite{Brown:2025xlo} for related classical constructions.}, consistent with
Rarita-Schwinger,
\begin{equation}
\boxed{
\text{Higher-Spin contact} = [\text{scalar block}] \times \prod_{\text{I}}  [\text{spin block}^{\text{(I)}} ]  } \,.
\end{equation}
However, constructing consistent, factorizing, interacting higher-spin
amplitudes is highly nontrivial and generally restricted by no-go theorems
unless embedded in string theory or extended frameworks. In contrast,
contact-level $n$-point amplitudes built from these double-copy states can often
be written down straightforwardly and used to explore EFT structures, soft
limits, and other consistency conditions, even in the absence of a fully
factorizing UV completion.

\subsection{Relation to the Traditional Double Copy}
\label{tradDCsec}
One of the key advantages of our modular double-copy framework is its ability to
generalize the traditional antisymmetric-adjoint double copy described by KLT,
BCJ, and CHY.
It is worth emphasizing that the modular block structure
presented here emerged from close examination of the internal algebraic
modularity, inspired by similar considerations for building antisymmetric-adjoint BCJ representations out of
mixtures of color and kinematics as per $Z$-theory
\cite{Broedel:2013aza,Carrasco:2016ldy,Mafra:2016mcc,Carrasco:2016ygv,Carrasco:2019yyn,Carrasco:2021ptp,Edison:2021ebi}.
This structure not only
underlies the familiar four-point double-copy amplitudes, but also enables
natural extensions to cases involving additional color tensors, such as the
symmetric~\cite{Carrasco:2023vjg} structure constants  $ d^{abc} $, and to theories with matter in the
fundamental representation—particularly for gauge groups that admit exchange
symmetry. By decoupling color and kinematics into manifestly compatible building
blocks, our approach offers a unifying language for organizing and generalizing
double-copy constructions beyond the traditional antisymmetric-adjoint-only framework.

The traditional KLT/BCJ/CHY double copy constructions already play well with
certain classes of contact terms.
At four points, for massless theories in the adjoint representation with color
weights $ f^{abc} $, a theory is said to be color-dual if its numerators satisfy
the same Jacobi identity as the color factors:
\begin{equation}
    c_s^{ff} = c_t^{ff} + c_u^{ff} \quad \Leftrightarrow \quad n_s = n_t + n_u\,.
\label{ckJacobis}
\end{equation}
When this holds, the full color-dressed amplitude takes the form
\begin{equation}
\mathcal{A} = \frac{c_s n_s}{s} + \frac{c_t n_t}{t} + \frac{c_u n_u}{u}\,.
\end{equation}
This amplitude can be recast by expressing  $ c_u^{ff} = c_s^{ff} - c_t^{ff} $ and $ n_u = n_s - n_t $:
\begin{align}
\mathcal{A}
&= \frac{c_s^{ff} n_s}{s} + \frac{c_t^{ff} n_t}{t} + \frac{(c_s^{ff} - c_t^{ff})(n_s - n_t)}{u} \\
&= \frac{(-s - t)t\, c_s^{ff} n_s + (-s - t)s\, c_t^{ff} n_t + (c_s^{ff} - c_t^{ff})(n_s - n_t)\, s t}{s t u} \\
&= -\frac{(c_s^{ff} t + c_t^{ff} s)(n_s t + n_t s)}{s t u} \\
&= -(c_s^{ff} t + c_t^{ff} s) \times \left[ \text{kinematic weight} \right] \times \sigma_3^{-1}\,,
\label{eq:color-factorized-amp}
\end{align}
where we have used momentum conservation $ s + t + u = 0 $, and defined the scalar permutation invariant
\begin{equation}
\sigma_3 = s t u\,.
\end{equation}
This form reveals that any color-dual four-point amplitude in such a theory is
proportional to the symmetric color structure  $ c_s t + c_t s $, multiplied by
a kinematic weight and a universal scalar factor.

If two theories $A$ and $B$ are color-dual, then their double copy is obtained
by replacing the color factors in one theory with the numerators of the other:
\begin{align}
\mathcal{A}^{(A) \otimes (B)}
&= \frac{n^{(A)}_s n^{(B)}_s}{s} + \frac{n^{(A)}_t n^{(B)}_t}{t} +
\frac{n^{(A)}_u n^{(B)}_u}{u} \\
&= -\frac{(n^{(A)}_s t + n^{(A)}_t s)(n^{(B)}_s t + n^{(B)}_t s)}{s t u}\,.
\label{eq:dc-amp}
\end{align}
If $A$ and $B$ are specifically two modular contact amplitudes, both proportional to
$ c_s t + c_t s$, then they can both be rearranged as
\begin{align}
\mathcal{A}^{(A)} &= (c_s^{ff} t + c_t^{ff} s) \times \left[ \text{kinematic weight}^{(A)} \right] \\
&= - (c_s^{ff} t + c_t^{ff} s) \times \left[ -\sigma_3 \times \text{kinematic weight}^{(A)} \right] \times \sigma_3^{-1} \\
\mathcal{A}^{(B)} &= (c_s^{ff} t + c_t^{ff} s) \times \left[ \text{kinematic weight}^{(B)} \right] \\
&= - (c_s^{ff} t + c_t^{ff} s) \times \left[ -\sigma_3 \times \text{kinematic weight}^{(B)} \right] \times \sigma_3^{-1}\,.
\end{align}
The quantity $-\sigma_3 \times [\text{kinematic weight}] $ can be interpreted as
the numerator factor which is color-dual to $ c_s^{ff} t + c_t^{ff} s$:
\begin{equation}
n_s t + n_t s = s t \, A(1234) = s u A(1243) = t u A(1423)\,,
\end{equation}
with $ n_u = n_s - n_t $. Thus, the full color-dressed contact amplitude for
either $ A $ or $ B $ can be written in terms of cubic graphs, absorbing contact
contributions into the numerators.
It follows that the double copy between two such contact amplitudes takes the form
\begin{equation}
\boxed{
\mathcal{A}^{A \otimes B} = \sigma_3 \times \left[ \text{kinematic weight}^{(\text{A})} \right] \times \left[ \text{kinematic weight}^{(\text{B})} \right]\,.
}
\label{tradDC}
\end{equation}

By recognizing when a modular contact amplitude is proportional to a color
structure of the form $ c_s^{ff} t + c_t^{ff} s $, we can identify the corresponding
kinematic blocks as arising from color-dual cubic numerators. This allows us to
interpret such amplitudes as conventional double copies, providing a useful
bridge between our flexible, exchange-parity-organized local construction and
the established cubic-graph framework. Doing so ensures compatibility with the
known color-dual web of theories, enables operator uplift in effective field
theory, and highlights the broader unifying structure underlying
color-kinematics duality that motivates our modular approach when
applied to gauge theory higher-derivative predictions.

Importantly, all of the important structures linking
\cref{eq:color-factorized-amp} to \cref{tradDC} live within our modular
framework.  The defining color structure $ c_s^{ff} t + c_t^{ff} s $ corresponds to one of
our manifestly symmetric color building blocks, as described in
\cref{fourPointAdjointAntisymmetricColorWeight}. Additionally, we note that $
\sigma_3 = s t u $ is spanned by our scalar building blocks (cf.
\cref{eq:sigma-3}). This makes it clear
that the traditional double-copy form, \cref{tradDC}, emerges from our general
double copy, \cref{genDC}, as a specific scalar-weight combination with  modular
kinematic blocks that independently form spin-consistent contacts when dressed
in the adjoint with $c_s^{ff} t + c_t^{ff} s$. While these building blocks may carry
different spins or structures on the left and right copies, their product
remains little-group covariant and gauge-invariant, and defines a valid
gravitational state as long as spin-statistics are respected. It should be
obvious that there is nothing particularly unique or canonical about
\cref{tradDC} -- the same higher derivative contact can appear from the
double-copy of many pairs of higher-derivative gauge contacts -- for example
with consistent scalar weights shuffled between the kinematic weights of (A) and
(B).  A virtue of KLT/BCJ/CHY not apparent when solely looking at
higher-derivative four-point contact terms is the ensured consistency of
factorization -- a challenge to be addressed within our framework as we move to
higher multiplicity.

\subsection{Our LEGOs are made of LEGOs}
\label{legosOnLegosOnLegos}
It should be noted that the primary fermionic building blocks we presented here are only prime with respect to each other.
If we admit purely bosonic modular blocks such as 2S+2V, or 3S+1V, then we see that indeed a few of our fermionic blocks are already double-copies.
For example:
\begin{align}
    {\spinBlockV{0,7}{-}{+} } &=    \underbrace{{ \tr(F_3,F_4) }}_{ {n^{(+|+)}_{2s2v,2}}}    \underbrace{\pbarp{}}_{  \spinBlockS{5}{\tzero}{+}} \\
    { \spinBlockV{0,9}{-}{+} } &=\underbrace{ { (k_1 - k_2) \cdot F_3 \cdot F_4 \cdot (k_1 - k_2) }}_{ {n^{(+|+)}_{2s2v,4}} }    \underbrace{\pbarp{}}_{  \spinBlockS{5}{\tzero}{+}} \\
      \spinBlockSV{0,8}{+} &=  \underbrace{ (k_1 - k_2) \cdot F_4 \cdot k_3}_{ {n^{(-|\text{NA})}_{3s1v,3}} }  \underbrace{\pbarp{}}_{  \spinBlockS{5}{\tzero}{+}}\,. 
\end{align}

\section{Examples}
\label{sec:examples}

We now provide a number of examples of explicit effective operators in various
theories and demonstrate how they are covered by our building blocks. We will
show that our modular blocks are particularly efficient when it comes to
promoting them to operators. In most cases, the operator promotion just involves
replacing the external states with their corresponding fields.

\subsection{Matching to SMEFT operators}
First, we make contact with SMEFT operators. It should be noted that some of the
operators arising from our spinor blocks show up in the classification of LEFT
operators but are absent in the list of SMEFT operators. The absence of such
operators can be accounted for by including SMEFT operators with a more
expansive particle content and matching them with LEFT operators at potentially
loop-level \cite{Jenkins:2017jig,Liao:2020zyx}. As an example, a higher
derivative tree-level contact term corresponding to a two fermion and two vector
interaction can be spanned by a loop level amplitude obtained by sewing together
a lower dimensional four fermion interaction with two fermion plus vector
interactions \cite{Liao:2020zyx}. We note that any spinor blocks that are
inconsistent with the product group symmetries of the standard model have to be
excluded when matching to SMEFT operators.

The interplay between Parity and electroweak interactions in the Standard Model
(and our lack of Parity-odd operators) means that exactly covering the relevant
four-point SMEFT operators with our current building blocks is not possible.
However, we do schematically cover the parity-even $SU(3)$ sector.  Because the
gauge group is complex, the fundamental color structures are drawn from
$\chh{\NA}{\NA,\pm}{0}$.

We will variously use Refs.~\cite{Alioli:2020kez,Murphy:2020rsh,Li:2020gnx,Murphy:2020cly} as
points of comparison, depending on which operator presentations are easiest to
match onto ours.

\subsubsection{Four fermions}
With four fermions all in the fundamental of $SU(3)$, we have access to the
octet $(T^{b})_{1}{}^{\bar{2}} (T^{b})_{3}{}^{\bar{4}}$ and singlet
$\delta_{1}{}^{\bar{2}} \delta_{3}{}^{\bar{4}}$ color structures from $\cnn{0}$.
We restrict the discussion in this subsection to operators in 4D.
\begin{itemize}
    \item[\textbf{Dimension 6}] We have a single independent spinor block at dimension 6, given by
    \begin{equation}
        \spinBlock{4\psi,1,6}{+}{+} = \pbarpexpr{\gamma^{\mu}} \,
        \pbarpgenexpr{3}{4}{\gamma_{\mu}} 
    \end{equation}

    This produces the only dimension 6 four-fermion SMEFT operator in the following way:
\begin{equation}\label{eq:FourFermions6Dexample}
O^{6}_{4\psi} = \pbarpexpr{ \gamma^{\mu} T^{a} } \, \pbarpgenexpr{3}{4}{ \gamma_{\mu} T^{a} } 
\leftrightarrow \cnn{} \times \left( \spinBlock{4\psi,1,6}{+}{+} = \pbarpexpr{ \gamma^{\mu} } \, \pbarpgenexpr{3}{4}{ \gamma_{\mu} } \right).
\end{equation}
\item[\textbf{Dimension 7}]
At dimension 7, we have two independent spinor blocks, given by
      \begin{align}
             \spinBlockF{0,7,r}{+}{+} = \pbarpexpr{ \kslashdiff{3}{4}  } \pbarpgenexpr{3}{4}{ }   \,, \\
          \spinBlockF{0,7,l}{+}{+} = \pbarpexpr{   } \pbarpgenexpr{3}{4}{\kslashdiff{1}{2} }  \,. 
        \end{align}
Their corresponding SMEFT operators are given by  
        \begin{align}
            \left( \psi_1  T^a \gamma^{\mu}\psi_2 \right) 
            \left( \psi_3 T^a \dherm_{\mu}\psi_4 \right)  
            &\leftrightarrow \cnn{} 
            \left( \spinBlockF{0,7,r}{+}{+} = \pbarpexpr{ \kslashdiff{3}{4}  } \pbarpgenexpr{3}{4}{ }  \right)  \,, \\
             \left( \psi_1 T^a \dherm_{\mu}\psi_2 \right) 
            \left( \psi_3  T^a \gamma^{\mu}\psi_4 \right) 
            &\leftrightarrow \cnn{} 
            \left(\spinBlockF{0,7,l}{+}{+} = \pbarpexpr{   } \pbarpgenexpr{3}{4}{\kslashdiff{1}{2} } \right)  \,. 
        \end{align}

\item[\textbf{Dimension 8}]
The spinor blocks at dimension $8$ can be constructed from the single
independent dimension 8 spinor block, or by multiplying our dimension 6 spinor
blocks by Mandelstams. As such, the space of spinor blocks at dimension 8 is
spanned by
\begin{align}
   \spinBlock{4\psi,1,8}{-}{-},\,  \spinBlock{4\psi,1,6}{+}{+}\otimes \{s,t\}. 
\end{align}
The operators we can build out of these three independent spinor blocks span the
space of all possible dimension 8 SMEFT operators after including the
appropriate particle content:
\begin{align}
\label{eq:FourFermions8Dexample1}
O^{8}_{1,4\psi} &= D^{\nu}(\pbarpexpr{ \gamma^{\mu} T^{a} }) \,D_{\nu} (\pbarpgenexpr{3}{4}{ \gamma_{\mu} T^{a} }) 
\nn
&\leftrightarrow \cnn{} \times \left( \spinBlock{4\psi,1,6}{+}{+} = \pbarpexpr{ \gamma^{\mu} } \, \pbarpgenexpr{3}{4}{ \gamma_{\mu} } \right) \times s
\end{align}

\begin{align}\label{eq:FourFermions8Dexample2}
O^{8}_{2,4\psi} &= (\pbarpexpr{ \dherm^{\nu}\gamma^{\mu} T^{a} }) \, (\pbarpgenexpr{3}{4}{ \dherm_{\mu} \gamma_{\nu} T^{a} }) 
\nn
&\leftrightarrow \cnn{} \times \left( \spinBlock{4\psi,1,8}{-}{-} = 
\pbarpexpr{ \kslashdiff{3}{4} } \, 
\pbarpgenexpr{3}{4}{ \kslashdiff{1}{2} } \right) \times 1
\end{align}

\begin{align}\label{eq:FourFermions8Dexample3}
O^{8}_{3,4\psi} &= (\pbarpexpr{ \gamma^{\mu} T^{a} \dherm^{\nu} }) \,
(\pbarpgenexpr{3}{4}{ \gamma_{\mu} T^{a} \dherm_{\nu} }) 
\nn
&\leftrightarrow \cnn{} \times \left( \spinBlock{4\psi,1,6}{+}{+} = \pbarpexpr{ \gamma^{\mu} } \, \pbarpgenexpr{3}{4}{ \gamma_{\mu} } \right) \times (t - u)
\end{align}
\end{itemize}

For the case of massless scalars in 4D, Fierz identities allow us to relate the operators in \cref{eq:FourFermions8Dexample2} and \cref{eq:FourFermions8Dexample3} to each other.

\subsubsection{Two fermions and two gluons}
For two vectors and two gluons charged in $SU(3)$, the color structures we have
access to are $T^b f^{b a_3 a_4} \in \cnm{0}$ and $\{\delta_1{}^{\bar{2}}
\delta^{a_3 a_4}, T^b d^{b a_3 a_4}\} \subset \cnp{0}$. If the vectors are both
photons, then we can only use $\delta_1{}^{\bar{2}} \in \cnp{0}$.  With two
distinct vectors, we have $(T^{a_4})_1{}^{\bar{2}} \in \cnn{0}$.
\begin{itemize}
    \item[\textbf{Dimension 7}]
        In 4D, our dimension 7 spinor blocks are over-complete: 
        $\spinBlockV{4,7}{-}{+}$ and $\spinBlockV{0,7}{-}{+}$ are
        degenerate when the fermions are massless. 
        Thus we take $\spinBlockV{0,7}{-}{+}$ and \spinBlockV{2,7}{+}{-} as our
        two independent blocks.

The spinor block $\spinBlockV{0,7}{-}{+}=\tr(F_3, F_4)\pbarpexpr{} $ can be
dressed with color in four different ways, leading to the following operators:
\begin{align}
    &\pbarpexpr{} G^{a}_{\mu \nu} G^{a\mu \nu}, & &d^{abc} \pbarpexpr{ T^a }G^{b}_{\mu \nu} G^{c\mu \nu}, \nn
    & \pbarpexpr{ T^a }F_{\mu \nu} G^{a\mu \nu}, & & \pbarpexpr{}F_{\mu \nu} F^{\mu \nu} .
\end{align}
Similarly, the spinor block $\spinBlockV{2,7}{+}{-} = \pbarpexpr{ \gamma^{\mu \nu}}
F_{3,\mu \rho} F_4 ^{\rho \nu}$ allows two different color dressings,
corresponding to the following operators:
\begin{align}
    f^{abc}\pbarpexpr{ T^a \gamma^{\mu \nu}} G ^b_{3\mu}{}^{\rho} G_{4\rho \nu}^{c}, 
       \hspace{1cm} 
       \pbarpexpr{ T^a \gamma^{\mu \nu}} G^a_{3\mu}{}^{\rho} F_{4\rho \nu}. 
\end{align}
These operators span all possible $\psi^2 X^2$ LEFT operators at dimension $7$
\cite{Liao:2020zyx}. They do not show up in the classification of the dimension
7 SMEFT operators as they lead to scalar fermion currents with zero hypercharge
\cite{Lehman:2014jma}. Moreover, the tree level amplitudes of these operators
can be matched to a loop level amplitude given by sewing together a dimension 6
4-fermion amplitude with two 2F+1V amplitudes.

    \item[\textbf{Dimension 8}]
        Ref.~\cite{Li:2020gnx} reports five two-quark two-glue operators at mass
        dimension 8.  Two of them involve Parity-violating terms, and thus
        cannot be covered by our basis\footnote{One of these operators, $i
            f^{abc} G^{a}_{\mu \nu} \tilde{G}^{b \nu}{}_{\lambda}
            \left(\bar{q} \gamma^{\lambda} \dherm^{\mu} T^c q \right)$, is
            almost exactly our $\cnm{0} \spinBlockV{3,8}{+}{-}$ except that ours
            would require an additional $\gamma^5$ in the spin contraction to
            cancel out the one generated by the Clifford duality relation on
            $\gamma^{\mu \nu \rho}$.}.  The other three are relatively obvious
            to decompose into our building blocks:
        \begin{align}
            i f^{abc} G^{a}_{\mu \nu} G^{b \nu}{}_{\lambda} \left(\bar{q} \gamma^{\lambda} \dherm^{\mu} T^c q \right) 
            &\leftrightarrow 
            \cnm{0} \spinBlockV{1,8}{-}{-}\\
            \left.\begin{array}{r}
                i d^{abc} G^{a}_{\mu \nu} G^{b \nu}{}_{\lambda} \left(\bar{q}
                \gamma^{\lambda} \dherm^{\mu} T^c q \right) \\
                i  G^{a}_{\mu \nu} G^{a \nu}{}_{\lambda} \left(\bar{q} \gamma^{\lambda} \dherm^{\mu} q \right) 
        \end{array}\right\}
            &\leftrightarrow 
            \cnp{0} \spinBlockV{1,8}{-}{+}\,.
        \end{align}
        Notably, for 4D massless particles $\spinBlockV{3,8}{+}{-}$ and
        $\spinBlockV{1,8}{+}{2}$ are degenerate with $\spinBlockV{1,8}{-}{-}$
        and $\spinBlockV{1,8}{-}{+}$, and there is no way to lift
        $\spinBlockV{i,7}{\pm}{\pm}$ to dimension $8$ with a scalar prefactor,
        so with those restrictions our construction $\chh{\NA}{\pm}{0} \otimes
        \spinBlockV{i,8}{\pm}{\pm}$ is exactly one-to-one with the pure-glue
        SMEFT terms.
\end{itemize}

\subsubsection{Two fermions, one gluon, one Higgs}
\begin{itemize}
    \item[\textbf{Dimension 6}]
        Ref.~\cite{Murphy:2020rsh} reports one Parity-even operator, which
        matches up nicely with our blocks
        \begin{align}
            (\pbarpgenexpr{}{}{ \sigma^{\mu \nu} T^a }) H G^{a}_{\mu \nu}
            \leftrightarrow \cnn{0} \spinBlockSV{2,6}{+}
        \end{align}
    \item[\textbf{Dimension 7}]    
        Our spinor blocks at dimension 7 lead to the following operators
        \begin{align}
             (\pbarpgenexpr{}{}{T^a \gamma_{\nu} }) (D_{\mu} \phi) G^{a\mu \nu}&\leftrightarrow \cnn{0} \spinBlockSV{1,7}{+} , \nn
            (\pbarpgenexpr{}{}{T^a \dherm_{\mu} \gamma_{\nu} }) \phi G^{a\mu \nu} &\leftrightarrow
            \cnn{0} \spinBlockSV{1,7}{-}.
        \end{align}

        These operators are excluded in the classification of dimension-7 SMEFT
        operators as they lead to vector fermionic currents with hypercharge
        $\pm \frac{1}{2}$ \cite{Lehman:2014jma}.

    \item[\textbf{Dimension 8}]
        Ref.~\cite{Li:2020gnx} reports 3 relevant dimension-8 operators, of
        which two are Parity-even.  They are schematically
        \begin{align}
            G^{a}_{\mu \nu} \left(\pbarpgenexpr{}{}{ T^a D^\mu }\right)D^\nu H
            &\leftrightarrow \cnn{0} \spinBlockSV{0,8}{+}\\
            G^{a}_{\mu \lambda} \left( \pbarpgenexpr{}{}{ \gamma^{\nu \lambda} T^a }\right) D^\mu D_\nu H
            & \leftrightarrow \cnn{0} \notag \\
            &\hspace{-2em} \times\left( -2\spinBlockSV{0,8}{+} +2(m_2-m_1)\spinBlockSV{1,7}{+}-\frac{1}{2}(s-m_3^2 )\spinBlockSV{2,6}{+} \right)
        \end{align}
\end{itemize}

\subsection{Maximal SYM 2F+2V}
The well-known dimension-8 counterterm for maximal SYM is \cite{Green:1983hw}
\begin{align}
    \mathcal{A}_4^{(1)} &\sim  
    d^{a_1 a_2 a_3 a_4}
    s\, t\, A^{\text{tree}}_{\text{YM}}(1,2,3,4)
    \label{eq:statree}\\
                        &\sim \left.\left[ \underbrace{c^{dd}_s}_{\cpp{0}} + \underbrace{( c^{dd}_t + c^{dd}_u)}_{\cpp{0}} \right]\right|_{N_c \leftrightarrow \infty} 
    \times s\, t\, A^{\text{tree}}_{\text{YM}}(1,2,3,4)
\end{align}
where the kinematic piece $s\, t\, A^{\text{tree}}$ has a two-fermion
two-gluon component.  Using the IncreasingTrees package
\cite{Edison:2020ehu} (or knowledge of Feynman rules), we see that
\cref{eq:statree} contains terms with both 1 and 3 $\gamma$ insertions, and
because it is a permutation ``invariant'' must already have the correct
transformation properties.  As such, we expect it to be decomposable into
the $(-|+)$ basis structures from \cref{sec:twoFtwoV}.  In fact, we find that,
up to normalization, it is exactly
\begin{equation}
    s\, t\, A^{\text{tree}}_{\text{YM}}(1_f,2_f,3_g,4_g) \propto
\spinBlockV{3;8}{-}{+} -
\spinBlockV{1;8}{-}{+}\,.
\end{equation}

\subsection{Scalar theories}
In the context of the double copy, \cref{sec:DC}, colored scalar effective field
theories are important for lifting gauge-interacting fermions to
gravitationally-interacting ones.  Below, we briefly discuss three examples,
demonstrating how they decompose into the
scalar kinematics blocks from \cref{sec:scalars} and the color blocks from
\cref{sec:color}, and explaining how they help organize higher-derivative
gravitational couplings.

\subsubsection{Minimally coupled adjoint scalar}
First we look at massless scalars interacting via a minimal gauge coupling.
The exchange process is described by a scattering amplitude of the following form:
\begin{equation}
\mathcal{A}^{D\phi} = \frac{c^{ff}_s n^{D\phi}_s}{s} +\frac{c^{ff}_t n^{D\phi}_t}{t} +\frac{c^{ff}_u n^{D\phi}_u}{u} \,,
\label{eq:cov-scalar}
\end{equation}
where the kinematic weights of the three channels are,
\begin{align}
    \label{covFreeScalarNums}
    n^{D\phi}_s = (t-u) , \qquad
    n^{D\phi}_t  = (s-u),\qquad
    n^{D\phi}_u = (t-s)\,.
\end{align}  
As these kinematic weights are manifestly antisymmetric around each vertex and
they satisfy a Jacobi relation in concordance with their color weights,
\begin{align}
    c^{ff}_s  &= c^{ff}_t + c^{ff}_u \\
    n^{D\phi}_s  &=  n^{D\phi}_t +  n^{D\phi}_u 
\end{align}
they are color dual. Therefore we can write the full amplitude in the form described
in \cref{tradDCsec}, entirely in terms of our LEGO blocks\,
\begin{align}
    \mathcal{A}^{D\phi}_4 
 &= - (c^{ff}_s t + c^{ff}_t s) (n^{D\phi}_s t + n^{D\phi}_t s) (s t u)^{-1} \\
 &= -  (c^{ff}_s t + c^{ff}_t s) ( 2 \sigma_2) ( \sigma_3)^{-1}   \\
 &\sim  \left[ \underbrace{ \left( c^{ff}_t - c^{ff}_u\right)}_{{\cpp{0}}} \underbrace{s}_{\ppp{2}} + \underbrace{c^{ff}_s}_{{\cmm{0}}} 
 \underbrace{ \left( t-u \right)}_{\pmm{2}} \right]
 \left[ 3 \!\!\! \underbrace{s^2}_{\ppp{4+0}} +\underbrace{(t-u)^{2}}_{\ppp{0 + 4}} \right] 
 \left[  \underbrace{s^3}_{\ppp{6+0}} -\underbrace{s (t-u)^2}_{\ppp{2+4}}
 \right]^{-1}\,.
\end{align}
The $\sigma_3=s\, t\, u$ in the denominator accounts for the factorization channels of the
propagating gluon. Note that while \cref{eq:cov-scalar,covFreeScalarNums} make
it clear that this is not a contact amplitude, it can be double-copied with
contact amplitudes to lift them to gravitational contacts \emph{without}
changing the external states.  Notably, doing so also shifts the mass dimension
by two.  As such, contact corrections that appear to descend from a
massive gauge mediator that are double-copied with \cref{eq:cov-scalar} will
produce a contact that appears to descend from a massive spin-two particle.

\subsubsection{NLSM pions}
Amusingly NLSM pions, which have only even point interactions, can be written in
terms of cubic graphs at 4-points by including an inverse propagator in their
kinematic numerator dressings,
\begin{equation}
\mathcal{A}^{\pi} = \frac{c^{ff}_s n^{\pi}_s}{s} +\frac{c^{ff}_t n^{\pi}_t}{t} +\frac{c^{ff}_u n^{\pi}_u}{u} \,,
\end{equation}
where the numerator weights are intimately related to those of the covariantized free scalars,
\begin{align}
\label{pionNumerators}
 n^{\pi}_s &= s(t-u)/3\,, \qquad
 n^{\pi}_t  = t(s-u)/3\,, \qquad
 n^{\pi}_u = u(t-s)/3\,.
\end{align}
Indeed one can see that $n^{\pi}_s \propto (n^{D\phi}_t)^2- (n^{D\phi}_u)^2$,
making pions in some sense a composition of covariantized free scalars that
preserves the duality between color and kinematics~\cite{Carrasco:2019yyn}.
Since the pion amplitude is a contact amplitude with no factorizable channels, it is entirely expressible in our
blocks:
\begin{align}
 \mathcal{A}^{\pi}_4 
 &= - (c^{ff}_s t + c^{ff}_t s) (n^{\pi}_s t + n^{\pi}_t s) (s t u)^{-1} \\
 &= -  (c^{ff}_s t + c^{ff}_t s) (- \sigma_3) ( \sigma_3)^{-1}   \\
   &= \text{\cref{fourPointAdjointAntisymmetricColorWeight}}\,.
\end{align}
We see an interesting feature: the full color dressed pion amplitude is the
ubiquitous permutation invariant color weight that appears 
in every four-point antisymmetric adjoint color-dual scattering amplitude,
$c^{ff}_s t + c^{ff}_t s$, as discussed in \cref{tradDCsec}.

\subsubsection{Capturing the rest of Z-theory (a bi-colored scalar effective field theory)}

$Z$-theory amplitudes allows us to understand tree-level string theory
amplitudes in terms of the double-copy of field-theory amplitudes.  Z-theory is
defined as the bi-colored  theory of all-order higher derivative corrections to
the bi-adjoint scalar theory that by double-copying with super Yang-Mills lifts
the field theory amplitude to the complete open-superstring amplitude.  
It can be expanded in terms of the string scale $\alpha'$, with
$\alpha' \to 0$ yielding the field-theory limit and higher derivative
corrections, or $\alpha' \to \infty$ probing intrinsically stringy operators.   
Schematically, it can be written as
\begin{align}
    [Z\text{-theory}]  &= [\text{scalar blocks}] \times [\text{color-blocks}] \times [\widetilde{\text{color-blocks}}] 
\end{align}
where one of the colors corresponds to Chan-Paton factors that appear in
open-string amplitudes and the other color is the usual antisymmetric adjoint
color -- which is stripped when double-copying with field theories to lift them
to string theories. 

At arbitrary multiplicity, Z-theory amplitudes are best
understood in terms of disk integrals.  However, the 4-point amplitude for
Z-theory has a simple closed form representation based on the Veneziano
amplitude \cite{Veneziano:1968yb}, which we will provide here then demonstrate how every
mass dimension of its $\alpha'$ expansion can be described in terms of our
blocks.  The closed form expression given in terms of Euler gamma functions is
as follows:
\begin{align}
\mathcal{A}^{Z}_4 &= -\frac{1}{s t u} \frac{\csc( \pi \alpha' s) \csc( \pi
\alpha' t ) \csc( \pi  \alpha' u)}{\Gamma(-s \alpha') \Gamma(-t \alpha') \Gamma(-u \alpha')} \times
 \alpha'^{-1} \left\{ z_s c_s   + z_t c_t + z_u c_u    \right.\notag \\ 
&\qquad  \left. + 2 \, \left [ \sin({ \pi \alpha' s})+ \sin({ \pi \alpha' t})+ \sin({ \pi \alpha' u}) \right] d^{a_1 a_2 a_3 a_4}  \right\} \times \left (\tilde{c}_s t + \tilde{c}_t s \right) \, ,
\end{align}
with $d^{abcd}$ the normalized permutation invariant sum over all distinct
color-traces, $z_s = \frac{ \pi^2}{\alpha'} (\sin(\pi \alpha' u) - \sin(\pi
\alpha' t))/3$ , $z_t = z_s |_{s\leftrightarrow t}$, and $z_u = z_s
|_{s\leftrightarrow u}$.  It should be clear that 
the amplitude can be organized as 
\begin{align}
    \mathcal{A}^Z_4 &=  - \frac{(\tilde{c}_s t + \tilde{c}_t s)}{s t u}  \times [\text{mixed kinematic-color block}]\,
\end{align}
where the mixture of Mandelstams and Chan-Paton factors in the [mixed
kinematic-color-block] satisfy permutation invariance.

We begin unpacking the mixed blocks by studying the $z_i c_i$ terms.
The $z_i$ are color-dual:
$z_s$ is manifestly antisymmetric under $t\leftrightarrow u$, and
the three channels satisfy Jacobi
equations in concordance with the color-weights,
\begin{equation} 
z_s = z_t + z_u\,.
\end{equation} 
The sum  $  z_s c_s   + z_t c_t + z_u c_u $ then can be
recognized as manifestly permutation invariant, so must be expressible
order-by-order in terms of our color and scalar building blocks.  Interestingly,
we only need two specific terms mixing color and kinematics, with the rest of
the behavior covered by an infinite series in kinematics only
\begin{multline}
    z_s c_s   + z_t c_t + z_u c_u =( c_s t + c_t s)  \frac{s (u-t) S_s +t (s-u) S_t +u (t-s) S_u}{(s - t) (s - u) (t - u) } \\
    -\frac{1}{3}[ \left( c_s s (t-u) + c_t t (s-u) + c_u u (t-s)  \right)  ] \frac{ (u-t) S_s +(s-u) S_t+(t-s) S_u}{(s - t) (s - u) (t - u) }\,,
\end{multline}
where $S_p = \frac{\pi^2}{\alpha'}\sin{\pi \alpha' p}$.  The two
mixed-color-kinematics directions are intimately related to objects we have already seen:
\begin{align}
    c^{ff}_s t + c^{ff}_t s 
    &= (c^{ff}_s n^{D\phi}_s+ c^{ff}_t  n^{D\phi}_t + c^{ff}_u  n^{D\phi}_u)/3  
    =  \text{\cref{fourPointAdjointAntisymmetricColorWeight}}\,,
\end{align}
\begin{align}
    \frac{1}{3}&[ \left( c_s^{ff} s (t-u) + c_t^{ff} t (s-u) + c_u^{ff} u (t-s)  \right)  ] 
    = c^{ff}_s n^{\pi}_s+ c^{ff}_t  n^{\pi}_t + c^{ff}_u  n^{\pi}_u \notag \\ 
    &=  \frac{1}{2} \underbrace{s (t-u)}_{\pmm{4}} \underbrace{c^{ff}_s}_{\cmm{0}}
    +\frac{1}{4}\left(- \underbrace{s^2}_{\ppp{4+0}} +\frac{1}{3} \underbrace{(t-u)^2}_{\ppp{0+4}}\right) 
    \left( \underbrace{c_t-c_u}_{\cpp{0}}\right)\,.
\end{align}
Because each of these color directions is manifestly permutation invariant, 
the series expansion of their kinematic coefficients in $\alpha'$ must
be expressible in terms of polynomials of $\sigma_2$ and $\sigma_3$ -- and thus
our scalar blocks via \cref{eq:sigma-2,eq:sigma-3}.
Finally, the permutation invariant scalar Veneziano factor can be rewritten as
\begin{align}
    \frac{\csc( \pi \alpha' s) \csc( \pi  \alpha' t ) \csc( \pi  \alpha' u)}{\Gamma(-s \alpha') \Gamma(-t \alpha') \Gamma(-u \alpha')}  &=
    -\frac{1}{\pi^3} \exp\left[ \sum_{n=2}^{\infty} (-1)^n \frac{ \zeta_n}{n}  (s^n+t^n+u^n) \alpha'^n \right] \,,
\end{align}
which by \cref{eq:sigma-n} is always order-by-order spanned by $\ppp{n}$.

Even though this is only a scalar amplitude, this tower of higher-derivative
operators mixing color and kinematics is absolutely non-trivial and is a nice
validation of how well our color and scalar blocks play together.   Of course
replacing the adjoint $(\tilde{c}^{ff}_s t + \tilde{c}^{ff}_t s)$ with any state
component of $s t A^{\text{SYM}}(1234)$ is exactly what one would do to execute
a traditional double-copy and results in the open superstring amplitude with
those external states.  For example replacing this tilded color-block with
$\spinBlockV{3;8}{-}{+} - \spinBlockV{1;8}{-}{+}$ yields two R-sector fermions
and two NS-sector vector components of the open-superstring vector multiplet.
Thorough analysis of double-copying other spin blocks with Z-theory is left for
the future.

\section{Conclusions}

The  modular framework introduced here provides a novel and systematic approach to constructing higher-derivative four-point contact interactions. The explicit $D$-dimensional nature of our kinematic building blocks ensures robust handling of loop-level structures, such as evanescent operators, critical for consistent EFT calculations and renormalization group evolution.  Furthermore, the demonstrated factorization into spin, color, and scalar polynomial components, where the latter systematically control the progression to arbitrary mass dimension, greatly simplifies the generation of complete operator bases for effective field theories.  In this proof of concept we specialized to four-points and arbitrary dimensions -- restricting ourselves to the spacetime parity even sector.  Spacetime parity-odd pieces fall perfectly in line with the above modular approach once one fixes to a particular dimension.  Of course the precise interface of  such expressions  with specific dimensional regularization schemes for chiral theories at loop level famously requires care. 

While we spent an entire paper talking about operators we did so in the language of amplitudes.  We note that mapping from contact amplitudes to quantum operators is as straightforward and mechanical as mapping from operators to amplitudes~\cite{deRoo:2003xv,Carrasco:2025ymt}.

The approach presented here not only offers a practical toolkit for phenomenological applications, such as building operator bases for SMEFT at dimension eight and (far) beyond, but also lays essential groundwork for exploring fundamental theoretical structures. The systematic construction of gauge theory contact terms in $D$ dimensions is a prerequisite for investigating their relationship to gravitational interactions via color-kinematics duality and the double-copy paradigm at the operator level especially as relates to known UV completions like string theory. The principles established here highlight a constructive path towards understanding the derivative expansions of gravitational effective actions from simpler gauge theory origins beyond the traditional anti-symmetric adjoint double-copy. We anticipate this framework will prove valuable in ongoing efforts to connect precision phenomenology with the fundamental theories of particle interactions and gravity.

\acknowledgments
We would  like to  thank members of the Amplitudes and Insight group at
Northwestern University, Yaxi (Sissi) Chen,  Nic Pavao,  Cong Shen, Theodore Wecker,
John Zhang, and Suna Zekiolğu for important conversations and related collaboration. 
We thank Supratim Das Bakshi for inspiring conversations on related
topics.
We thank Zhe Ren for discussion and feedback on the manuscript.
We especially thank Francis Pretriello for warm encouragement and helpful comments on an earlier draft.  
This work was supported by a grant from the Simons Foundation International
[SFI-MPS-SSRFA-00012751, AE].
This work was also supported in part by the DOE under contract DE-SC0015910, and by Northwestern University via the
Amplitudes and Insight Group, Department of Physics and Astronomy, and
Weinberg College of Arts and Sciences.
This research was supported in part through the computational resources and staff contributions provided for the Quest high performance computing facility at Northwestern University which is jointly supported by the Office of the Provost, the Office for Research, and Northwestern University Information Technology.

\appendix
\crefalias{section}{appendix}

\section{Three-point Fermionic LEGOs}
\label{app:3pt}
\newcommand{\Fone}{F_{1}^{\mu\nu}}
\newcommand{\Ftwo}{F_{2}^{\mu\nu}}
\newcommand{\vSFF}{\mathfrak{v}^{(+)}_{S;FF}}
\newcommand{\vSFFtil}{\tilde{\mathfrak{v}}^{(+)}_{S;FF}} 

\paragraph{Conventions.}
We consider $p_1{+}p_2{+}p_3=0$ with $p_i^2=m_i^2$. As Mandelstam invariants
reduce to masses, kinematic scalars are just mass monomials. Exchange parity
refers to $1\leftrightarrow 2$.  We employ the Majorana flip convention defined
in \cref{sec:spinors}, where the parity of a $k$-gamma structure, $t_k$, is used
for labeling.  Recall that we use the $D=4$ choice where $t_1=+1$ and  $t_0=-1$.

\subsection{Scalar blocks}
\paragraph{Even under $1\leftrightarrow 2$ $(+)$:}
\[
\mathcal{P}_D^{(+)}= (m_1+m_2)^{a_1}\, (m_1 m_2)^{a_2}\,m_3^{a_3}\,,\qquad
 D=a_1+a_3+2a_2.
\]

\paragraph{Odd under $1\leftrightarrow 2$ $(-)$:}
\[
\mathcal{P}_D^{(-)}=(m_1{-}m_2)\mathcal{P}_{(D-1)}^{(+)}
\]

\subsection{Color blocks at 3pt}
Only rank-3 tensors appear:
\[
(+):\ d^{a_1 a_2 a_3},\qquad
(-):\ f^{a_1 a_2 a_3},\qquad
(-) |  (\text{NA}):\ (T^{a_3})_{i j}\,,
\]
with the fundamental case ``NA'' if the rep is complex (no well-defined exchange parity).

\subsection{Spin blocks with Fermions}
\paragraph{Two fermions + one scalar.}
All that is available is Yukawa which is odd in fermion exchange.
\begin{align*}
\text{Odd }(-):\quad
& n^{(-)}_{s,5}=\bigl(\bar\psi_1 \psi_2\bigr)\,,
\end{align*}

\paragraph{Two fermions + one massless vector.}
We keep one photon/gluon on leg~3 with polarization $\pol_q^\mu$ and linearized field strength
$F^{\mu\nu}\!=p_3^{[\mu}\pol_3^{\nu]}$, so every term is linear in $\pol_q$ and gauge invariant.
\begin{align*}
\text{Even }(+):\quad
& n^{(+)}_{v,5}=\bigl(\bar\psi_1\slashed{\pol_3} \psi_2\bigr) \,   \qquad (\text{if }m_1=m_2), \\
& n^{(+)}_{v,6}=\pbarpexpr{\slashed{F}_3}\,.
\end{align*}
Higher-derivative towers are obtained by multiplying by the scalar mass blocks
$\mathcal{P}^{(\pm|+)}_{D}$.

In the equal-mass case, we can instead make the choice
\begin{equation}
    n^{(+)}_{v,6b} = (\pbarp{}) (p_1 - p_2) \cdot \pol_3 =
    \pbarpexpr{\slashed{F}_3} - (m_1 + m_2) \pbarpe{3}\,.
\end{equation}
This choice is interesting because, similar to the situation in
\cref{legosOnLegosOnLegos}, $n^{(+)}_{s,6b}$ can be understood as composite if
we admit 2s+1v building blocks
\begin{equation}
 n^{(+)}_{s,6b}=\underbrace{\bigl(\bar\psi_1 \psi_2\bigr) }_{{ n^{(-)}_{s,5} }}\, \underbrace{ (p_1{-}p_2)\!\cdot\!\pol_3 }_{ { n^{(-)}_{2s1v,1} } } \,.
\end{equation}

\section{Four fermion spinor blocks in 4D}
\label{sec:4F4D}
While the entirety of the basis we listed in \cref{sec:fourFermions} is required
to span all of the on-shell spinor blocks in arbitrary spacetime dimensions, it
is over-complete in 4D. There, one can use Fierz identities to eliminate any
summation over $\Gamma^A$s, potentially at the cost of including the charge
conjugated fields in our on-shell basis. We provide two examples of such
degeneracies in 4D. First consider the basis element
$\left(\pbarpexpr{\gamma^{\mu} \gamma^{\nu} } \right) \left(
\pbarpgenexpr{3}{4}{\gamma_{\mu} \gamma_{\nu} } \right)$. When the helicity
configuration is $(1^+ 2^+ 3^+ 4^+)$, this expression takes the form
\begin{align}
    (\pbarpexpr{\gamma^{\mu} \gamma^{\nu} })( \pbarpgenexpr{3}{4}{\gamma_{\mu} \gamma_{\nu} } )\sim [1|\gamma^{\mu} \gamma^{\nu}| 2] [3|\gamma_{\mu}\gamma_{\nu}| 4] \sim [12][34] \sim \pbarpexpr{ } \pbarpgenexpr{3}{4}{ }.
\end{align}
This element is already spanned by a permutation of the spinor block given by
$\pbarpexpr{ } \pbarpgenexpr{3}{4}{}$. Second, consider the spinor
block $(\pbarpexpr{\gamma^{\mu} \gamma^{\nu}\gamma^{\rho} })(
\pbarpgenexpr{3}{4}{\gamma_{\mu} \gamma_{\nu}\gamma_{\rho}  } )$. When the helicity
configuration is $(1^+ 2^+ 3^+ 4^+)$, this expression reduces to
\begin{align}
    (\pbarpexpr{\gamma^{\mu} \gamma^{\nu}\gamma^{\rho} })( \pbarpgenexpr{3}{4}{\gamma_{\mu} \gamma_{\nu}\gamma_{\rho}  } )
    &\sim  [1|\gamma^{\mu}\gamma^{\nu}\gamma^{\rho}| 2 \ra [3|\gamma_{\mu}\gamma_{\nu}\gamma_{\rho}| 4 \ra \nn
    &\sim [13]\la 24 \ra \sim (\overline{\psi}_1 \overline{\psi}_3 ^C )(\psi_2 \psi_4 ^C).
    \label{eq:fierz-four}
\end{align}
We can eliminate the need for introducing charge conjugated fields by
reintroducing a summation over a single gamma matrix:
\begin{align}
    (\pbarpexpr{\gamma^{\mu} \gamma^{\nu}\gamma^{\rho} })( \pbarpgenexpr{3}{4}{\gamma_{\mu} \gamma_{\nu}\gamma_{\rho}  } ) 
    \sim (\overline{\psi}_1 \overline{\psi}_2 ^C )(\psi_2 \psi_4 ^C) 
    \sim (\pbarpexpr{\gamma^{\mu} })( \pbarpgenexpr{3}{4}{\gamma_{\mu}  } ) .
    \label{eq:fierz-four-2}
\end{align}

We list the 4D representations of the reduced basis of spinor blocks for 4
fermion interactions that we outlined in \cref{sec:fourFermions} below. We use
the parity signatures corresponding with the Majorana flip condition that are
consistent with the standard 4D specific convention outlined at the beginning of
\cref{sec:spinors}.

\begin{itemize}
    \item[$(+|+)$] Even-Even:  
        \begin{align}
            \spinBlockF{0,7,r}{+}{+} &= \pbarpexpr{ \kslashdiff{3}{4}  } \pbarpgenexpr{3}{4}{ } \,, \\
            \spinBlockF{0,7,l}{+}{+} &= \pbarpexpr{   } \pbarpgenexpr{3}{4}{\kslashdiff{1}{2} } \,, \\
            \spinBlockF{1,6}{\tone}{\tone} &= \pbarpexpr{ \gamma^\mu }\
            \pbarpgenexpr{3}{4}{ \gamma_\mu } \,. 
        \end{align}

    \item[$(-|+)$] Odd-Even: None

    \item[$(+|-)$] Even-Odd: None

    \item[$(-|-)$] Odd-Odd:
        \begin{align}
            \spinBlockF{0,8}{\tzero}{\tzero} &= \pbarpexpr{
            \kslashdiff{3}{4} }\ \pbarpgenexpr{3}{4}{ \kslashdiff{1}{2}  } \,.
        \end{align}
\end{itemize}

\section{Spinor helicity expressions for the spin building blocks }
\label{sec:sh-proj}
In this section, we present the spinor helicity expressions for the spin
building blocks corresponding to 2 fermions + 2 vectors and 2 fermions +1 scalar
+ 1 vector. The expressions we list below hold true for both massive and
massless fermions. We use the notation from refs.
\cite{Dixon:1996wi,Gelis:2019yfm} and take
\begin{align}
    \epsilon_{+} (p,q) &= \frac{\la q| \overline{\sigma}^{\mu}|p]}{\sqrt{2} \la qp \ra}, \nn
    \epsilon_{-} (p,q) &= -\frac{[ q| \sigma^{\mu}|p\ra}{\sqrt{2} [qp]}.
\end{align}
We don't choose any particular helicity basis for the external spinors and leave
them arbitrary, so that they may safely be chosen either massive or massless.
The vast majority of algebraic simplifications due to spinor helicity in this
situation are due to the manifest gauge invariance of the massless vectors, so
leaving the spinors themselves unprojected favors flexibility over a few final
simplifications.

\subsection{2F+2V}
\begin{itemize}
    \item[$(+|+)$] Even-Even
\begin{itemize}
\item[N/A] 
\end{itemize}

    \item[$(-|+)$] Odd-Even
        \begin{itemize}
            \item[$\spinBlockV{0,7}{-}{+}$]$=\tr(F_3 \cdot F_4) (\pbarpexpr{ }) $
                \begin{align}
                    n(1 2 3^{\pm} 4^{\mp}) &=0, \nn
                    n(1 2 3^+ 4^+) &=-[34]^2 \left(\pbarpexpr{ } \right), \nn
                     n(1 2 3^- 4^-) &=-\la 34 \ra ^2 \left(\pbarpexpr{ } \right).
                \end{align}

            \item[$\spinBlockV{0,9}{-}{+}$]$ =(k_1-k_2) \cdot F_3 \cdot F_4 \cdot (k_1-k_2) (\pbarpexpr{ })$
                \begin{align}
                    n(1 2 3^+ 4^+) &= -\frac{[34]^2}{4}  (k_1 - k_2)^2  \left(\pbarpexpr{ } \right) ,  \nn
                    n(1 2 3^- 4^-) &= -\frac{\la 34 \ra^2}{4}  (k_1 - k_2)^2  \left(\pbarpexpr{ } \right) ,  \nn
                    n(123^+ 4^-) &= \la 4|1|3]^2 \left(\pbarpexpr{ } \right),  \nn
                    n(123^- 4^+) &= [ 4|1|3 \ra^2 \left(\pbarpexpr{ } \right).
                \end{align}

            \item[$\spinBlockV{1,8}{-}{+}$]$ =((k_1-k_2)\cdot F_4 \cdot F_3) ^\mu \pbarpexpr{ \gamma_{\mu} } + (3 \leftrightarrow 4) $
                \begin{align}
                    n(1 2 3^+ 4^+)&=\frac{[34]^2}{2}(m_1+m_2)\pbarpexpr{ }, \nn
                                  &\text{This vanishes when the two fermions are massless}. \nn
                                  n(1 2 3^+ 4^-) & =2 \la 4|\slashed{1}|3]  \pbarpexpr{ \left(|3]\la 4| + |4 \ra [3|  \right)} , \nn
                                  n(1 2 3^- 4^+) & =2 \la 3|\slashed{1}|4]  \pbarpexpr{ \left(|4]\la 3| + |3 \ra [4|  \right)}.
                              \end{align}

                          \item[$\spinBlockV{3,8}{-}{+}$]$ = \psi_1 \gamma^{\mu \nu \rho} \psi_2 (F_3 \cdot F_4)_{\mu \nu} (k_3-k_4)_{\rho}$
                              \begin{align}
                                  n(123^{\pm} 4^{\mp})&=0, \nn
                                  n(1  2 3^+ 4^+) &=\frac{1}{2}[34]^2 \overline{\psi}_1 (|3 \ra [3|-|3] \la 3|+|4 \ra [4|-|4] \la 4|)\psi_2.
                              \end{align}
                              This vanishes when the two fermions are massless.

                          \item[$\spinBlockV{4,7}{-}{+}$]$ = \psi_1 \slashed{F}_3 \slashed{F}_4 \psi_2 + (3\leftrightarrow 4)$
                              \begin{align}
                                  n(123^{\pm} 4^{\mp})&=0, \nn
                                  n(1 2 3^+ 4^+) &=2[34] \pbarpexpr{ \left(|3][4|-|4][3|\right) }, \nn
                                   n(1 2 3^- 4^-) &=2\la 34 \ra \pbarpexpr{ \left(|3 \ra \la 4|-|4 \ra \la 3|\right) }.
                              \end{align}
                      \end{itemize}

                  \item[$(+|-)$] Even-Odd
                      \begin{itemize}
                          \item[$\spinBlockV{1,8}{+}{-}$]$ =\tr(F_3 \cdot F_4)\pbarpexpr{ (\slashed{k}_3 -\slashed{k}_4) }$
                              \begin{align}
                                  n(123^{\pm}4^{\mp})&=0, \nn
                                  n(1 2 3^+ 4^+) &=[34]^2  \pbarpexpr{ \left( \slashed{k}_4-\slashed{k}_3\right) }, \nn
                                   n(1 2 3^- 4^-) &=\la 34 \ra ^2  \pbarpexpr{ \left( \slashed{k}_4-\slashed{k}_3\right) }.
                              \end{align}

                          \item[$\spinBlockV{2,7}{+}{-}$]$ = \pbarpexpr{ \gamma^{\mu \nu} } (F_3 \cdot F_4)_{\mu \nu}$
                              \begin{align}
                                  n(1 2 3^+ 4^-)&=0, \nn
                                  n(1 23^+ 4^+)&=[34] \pbarpexpr{\left( |3 ][4|+|4][3|\right)}.
                              \end{align}

                          \item[$\spinBlockV{3,8}{+}{-}$]$ =\pbarpexpr{ \gamma^{\mu \nu \rho} } F_{3\mu \nu} F_{4 \rho \sigma} (k_1 - k_2)^{\sigma} - (3 \leftrightarrow 4)$
                              \begin{align}
                                  n(12 3^+ 4^+) &=  2[34]^2 \pbarpexpr{ \left(|3]\la 3| - |4]\la 4| \right)} \nn
                                  &\hspace{1cm}- 4(m_1+m_2) [34]\pbarpexpr{ |3][4|} -[34]^2(m_1+m_2)\pbarpexpr{} ,\nn
                                  n(123^+ 4^-)&= 4[3|1|4\ra  \pbarpexpr{\left( |3]\la 4| - |4\ra [3| \right)}.
                              \end{align}

                      \end{itemize}

                  \item[$(-|-)$] Odd-Odd
                      \begin{itemize}
                          \item[$\spinBlockV{1,8}{-}{-}$]$=((k_1-k_2)\cdot F_4
                              \cdot F_3)_\mu \pbarpexpr{ \gamma^{\mu} } - (3 \leftrightarrow 4) $
                              \begin{align}
                                  n(1 2 3^+ 4^-)&= 0, \nn
                                  n(1 2 3^+ 4^+)&=(m_1-m_2)[34] \pbarpexpr{ \left(|3][4|+|4][3|  \right)} ,\nn
                                                &\hspace{1cm}+ \frac{[34]^2}{2} \pbarpexpr{ \left(|3]\la 3|-|4]\la 4| - |3 \ra [3|+|4 \ra [4|\right) }.
                              \end{align}

                          \item[$\spinBlockV{2,9}{-}{-}$]$= (\pbarpexpr{ \slashed{F}_3  }) (k_1-k_2)\cdot F_4  \cdot k_3 - (3 \leftrightarrow 4)  $
                              \begin{align}
                                  n(1 2 3^+ 4^+) &=\pbarpexpr{ \left( |3][3| [4|\slashed{3} \slashed{1}|4] - (3 \leftrightarrow 4) \right) }, \nn
                                                 &\text{the expression above vanishes when all particles are massless}\nn
                                  n(1^+ 2^+ 3^+ 4^-) &=[3|\slashed{1}|4 \ra
                                  \pbarpexpr{\left( \la 3 4 \ra |3][3| + [34] |4 \ra \la 4| \right)}.
                              \end{align}
                      \end{itemize}
              \end{itemize}

              \subsection{2F+1S+1V}
              In this particular subsection, we take the scalar particle to be massless for
              the spinor helicity expressions so it can serve as the reference momenta for $\epsilon_4$.
              \begin{itemize}
                  \item[$\spinBlockSV{0,8}{+}$]$=(k_1-k_2)\cdot F_4 \cdot k_3 \pbarp{}$ 
                      \begin{align}
                          n(1 2 3 4^+)&=\frac{1}{\sqrt{2}}[34]\la 3 |\slashed{1} | 4] \pbarp{}.
                      \end{align}

                  \item[$\spinBlockSV{1,7}{-}$]$=(k_1-k_2)_{\mu} \pbarpf{4}$
                      \begin{align}
                          n(1 2 3 4^+)=&\sqrt{2}(m_1-m_2)\pbarpexpr{|4][4|} +
                      \frac{[34]}{\sqrt{2}} \pbarpexpr{\left(|4]\la 3| -| 3
                      \ra[4| \right)}.
                      \end{align}

                  \item[$\spinBlockSV{1,7}{+}$]$=k_{3\mu} \pbarpf{4}$
                      \begin{align}
                          n(1 2 3 4^+)&=-\frac{[34]}{\sqrt{2}}
                          \pbarpexpr{\left(|4]\la 3|+|3 \ra [4|\right)}.
                      \end{align}

                  \item[$\spinBlockSV{2,6}{+}$]$=  \pbarpexpr{\slashed{F}_4}$
                      \begin{align}
                          n(1 2 3 4^+)&=-\sqrt{2}\, \pbarpexpr{|4][4|}.
                      \end{align}
              \end{itemize}

              \bibliographystyle{JHEP}
              \bibliography{references}

\end{document}